\documentclass[10pt,journal,compsoc]{IEEEtran}
\ifCLASSOPTIONcompsoc
  \usepackage[nocompress]{cite}
\else
  \usepackage{cite}
\fi
\ifCLASSINFOpdf
\else
\fi
\usepackage{amsmath,amssymb,amsfonts}
\newtheorem{thm}{Theorem}

\newtheorem{defn}[thm]{Definition} 

\usepackage{array}
\usepackage[ruled,vlined]{algorithm2e}
\usepackage{algorithmic}
\usepackage{graphicx}
\usepackage{subfigure} 
\usepackage{textcomp}
\usepackage{color} 
\usepackage{soul}
\usepackage{url}
\usepackage{multirow}
\usepackage{amsfonts}
\usepackage{xcolor}
\usepackage{enumerate}
\definecolor{blue}{RGB}{0,0,0}
\definecolor{red}{RGB}{223,58,45}

\begin{document}
%
\title{Joint Link Scheduling and Power Allocation in Imperfect and Energy-Constrained Underwater Wireless Sensor Networks}
%
%
%
%

\author{Tong Zhang,
	Yu Gou,
        Jun Liu,~\IEEEmembership{Member,~IEEE},
        Shanshan Song,
        Tingting~Yang,~\IEEEmembership{Member,~IEEE},
        and~Jun-Hong~Cui
\IEEEcompsocitemizethanks{
\IEEEcompsocthanksitem Tong Zhang and Yu Gou are with Beihang Ningbo Innovation Research Institute, Beihang University, Ningbo, China, 315800, School of Electronic and Information Engineering, Beihang University, Beijing, China, 100191, and also with the College of Computer Science and Technology, Jilin University, Changchun, China, 130012. E-mail: \{tongzhang18, gouyu18\}@mails.jlu.edu.cn.
\IEEEcompsocthanksitem Jun Liu is with the School of Electronic and Information Engineering, Beihang University, Beijing, China, 100191. E-mail: liujun2019@buaa.edu.cn.
\IEEEcompsocthanksitem Shanshan Song is with the College of Computer Science and Technology, Jilin University, Changchun, China, 130012. 
\IEEEcompsocthanksitem Tingting Yang is with Department of Network Intelligence, Peng Cheng Laboratory, Shenzhen, China, and also with Navigation College, Dalian Maritime University, Dalian, China. E-mail: yangtingting820523@163.com.
\IEEEcompsocthanksitem Jun-Hong Cui is with Shenzhen Institute for Advanced Study, UESTC, Shenzhen, China, and also with the College of Computer Science and Technology, Jilin University, Changchun, China.
}
\thanks{(Corresponding author: Jun Liu.)}
}

%
%

\markboth{Journal of \LaTeX\ Class Files,~Vol.~14, No.~8, August~2015}%
{Shell \MakeLowercase{\textit{et al.}}: Bare Demo of IEEEtran.cls for Computer Society Journals}
%



\IEEEtitleabstractindextext{%
\begin{abstract}
Underwater wireless sensor networks (UWSNs) stand as promising technologies facilitating diverse underwater applications. However, the major design issues of the considered system are the severely limited energy supply and unexpected node malfunctions. This paper aims to provide fair, efficient, and reliable (FER) communication to the imperfect and energy-constrained UWSNs (IC-UWSNs). Therefore, we formulate a FER-communication optimization problem (FERCOP) and propose ICRL-JSA to solve the formulated problem. ICRL-JSA is a deep multi-agent reinforcement learning (MARL)-based optimizer for IC-UWSNs through joint link scheduling and power allocation, which automatically learns scheduling algorithms without human intervention. However, conventional RL methods are unable to address the challenges posed by underwater environments and IC-UWSNs. To construct ICRL-JSA, we integrate deep Q-network into IC-UWSNs and propose an advanced training mechanism to deal with complex acoustic channels, limited energy supplies, and unexpected node malfunctions. Simulation results demonstrate the superiority of the proposed ICRL-JSA scheme with an advanced training mechanism compared to various benchmark algorithms.
\end{abstract}

}

\maketitle

\IEEEdisplaynontitleabstractindextext

%
\IEEEpeerreviewmaketitle

\IEEEraisesectionheading{\section{Introduction}\label{sec:intro}}

\IEEEPARstart{C}{onsisting} of nodes with sensing and communication capabilities, underwater wireless sensor networks (UWSNs) have emerged as a research hotspot due to their extensive applications. For example, UWSNs can be utilized for tsunami disaster reduction through the implementation of tsunami early warning systems, which leverage seismic activity detection and analysis to provide timely tsunami warnings to officials and coastal communities \cite{satake2014advances}. Furthermore, UWSNs find utility in monitoring anomalous events in underwater oil and gas pipelines, with the ability to relay diagnostic information to remote control stations \cite{aalsalem2018wireless}. The defense and security industries also benefit from UWSNs, facilitating communication among submarines, surface gateways, and underwater combat platforms \cite{xu2016digital}.

Prior to the practical implementation of UWSNs in the aforementioned scenarios, researchers face several challenges that need to be addressed. First, acoustic channels are considered one of the most challenging communication media due to limited bandwidth, long propagation delays, and time-varying channel conditions \cite{stojanovic2009underwater}. This leads to lower-than-desired delivery ratios in underwater acoustic communications and diminished UWSN capacity. Second, the reliance on batteries to power most underwater nodes introduces significant constraints, given the high cost and inherent challenges associated with replacing or recharging batteries in underwater environments \cite{gou2022DMPM}. Tthe lifetime of UWSNs is limited by the availability of energy supplies.

Additionally, existing network optimization solutions always assume that the underwater nodes can function normally all the time, which is usually not the case. In real-world applications, the mismatch between simulation and reality can lead to severe performance degradation of the optimization algorithms. Consequently, UWSNs exhibit dual characteristics: 1) \textit{imperfect}, given the complexity of acoustic channels and the the potential for unexpected malfunctions in underwater nodes, and 2) \textit{energy-constrained}. This paper mainly investigates the performance of imperfect and energy-constrained UWSNs (IC-UWSNs).

Due to spatial separation, multiple underwater nodes can transmit simultaneously to enhance spatial reuse and network capacity, provided that one transmission is not destructively interfered with another \cite{gupta2000capacity}. In multi-user systems, the optimization methods typically include communication fairness as an additional performance criterion to prevent unfair resource allocation that may lead to flow starvation \cite{sathiaseelan2007multimedia}. However, achieving communication fairness often involves trade-offs with network capacity \cite{diamant2016leveraging}. Providing additional transmission opportunities to nodes with superior performance can boost network capacity but may result in delivery delays for other nodes \cite{gou2022DMPM}. Conversely, ensuring fair transmission opportunities for all communication links, even those disadvantaged by deployments or interference from concurrent communications, may compromise network capacity and reliability \cite{gou2021achieving}. Therefore, there is a pressing need for an appropriate resource allocation scheme that simultaneously achieves fair, efficient, and reliable communication in IC-UWSNs.

Link scheduling (LS) and power allocation (PA) stand out as effective strategies for underwater resource management when optimizing spatial reuse, communication fairness, and communication quality \cite{zhang2021udarmf}. In UWSNs, characterized by unstable communication environments and functioning as typical multi-user systems, relying solely on PA algorithms proves insufficient to eliminate strong inter-node interference. Similarly, LS algorithms alone cannot guarantee communication quality \cite{elbatt2004joint}. The necessity for joint link scheduling and power allocation (JSA) schemes becomes evident to optimize spatial reuse, communication fairness, and reduce energy consumption \cite{cao2018machine}. Moreover, manual management of underwater nodes encounters significant challenges due to the complexity of underwater environment and sheer scale of UWSNs. Hence, it is necessary to control the devices in a distributed manner with minimal human intervention \cite{park2016learning}. Recent successes in employing reinforcement learning (RL) to adaptively address the dynamic performance optimization problems in UWSNs are noteworthy \cite{naderializadeh2021resource}. However, conventional RL methods fall short in addressing challenges specific to underwater environments and IC-UWSNs, particularly regarding the impact of node malfunctions on cooperative RL algorithms.

In this paper, a deep multi-agent reinforcement learning (MARL)-based joint link scheduling and power allocation scheme for IC-UWSNs, termed as ICRL-JSA, has been developed to address the aforementioned challenges and design issues. Nodes employing the ICRL-JSA scheme are referred to as ICRL-JSA nodes or intelligent nodes. ICRL-JSA serves as a general scheme for learning JSA policies for FERCOP in IC-UWSNs. We consider a distributed and cooperative ICRL-JSA implementation, where each ICRL-JSA node individually determines its transmission parameters, without relying on global information or engaging in negotiations with other nodes. To adapt conventional RL methods for IC-UWSNs, we modify the training procedure by early termination of training episodes based on energy constraints. Additionally, we present an advanced training mechanism allowing the adjustment of training difficulty during the training process. With these design considerations, ICRL-JSA is capable of initially learning to solve FERCOP in perfect UWSNs and then graduate to the more challenging IC-UWSNs scenarios.

Overall, the main contributions of this paper are summarized as follows:
\vspace{-1em}
\begin{itemize}
\item We summarize the main challenges and constraints in imperfect and energy-constrained UWSNs (IC-UWSNs). In particular, we formulate the fair, efficient, and reliable communication optimization problem (FERCOP) to explore the optimal trade-off design between network capacity and communication fairness, while enhancing network reliability through the augmentation of the delivery ratio.
\item We present ICRL-JSA, a deep MARL-based optimizer for IC-UWSNs, which solves the formulated problem by jointly scheduling communication links and allocating appropriate transmit power to scheduled nodes. The ICRL-JSA nodes decide their transmission parameters solely based on local observations without requiring global information or negotiating with other nodes, which imposes no communication overhead on the system.
\item We adapt the deep Q-network to IC-UWSNs, termed as IC-DQN, by integrating unexpected node malfunctions and limited energy supplies. IC-DQN terminates network service when the remaining energy of the node falls below its energy threshold to satisfy the energy constraint of FERCOP. Furthermore, we propose an advanced training mechanism that generates imperfect training environments for ICRL-JSA agents in response to model performance. The ICRL-JSA agents are always trained in a hortative environment and are constantly encouraged to improve.
\item Simulation results show the benefits of the proposed ICRL-JSA for solving the formulated FERCOP under different network density and communication scenarios. In addition, the simulation results also validate the effectiveness and necessity of ICRL-JSA's design components.
\end{itemize}

The remainder of the paper is organized as follows. In Section \ref{sec:rw}, we review the related work on link scheduling and power allocation algorithms. We describe the system model and formulate the problem in Section \ref{sec:sys}. The ICRL-JSA scheme is presented in Section \ref{sec:method}. Section \ref{sec:pe} gives the evaluation results. Finally, Section \ref{sec:con} concludes the paper.


\vspace{-1em}
\section{Related Work}\label{sec:rw}

The underwater acoustic channel is a complex communication medium, which is featured by long propagation delay, limited available bandwidth, and poor channel utilization. Due to the extensive applications of UWSNs, the optimization of their performance has garnered growing interest within both academic and industrial communities. Link scheduling and power allocation techniques have been extensively employed for the optimization of UWSNs. In \cite{gorma2019adaptive}, an adaptive and robust TDMA-based link scheduling algorithm is devised to optimize channel utilization and transmission delay. This centralized solution employs a central controller that combines round-robin free allocation with a demand assignment scheme. Considering the acoustic channel conditions and the limited battery capacity of the underwater nodes, Wang \textit{et al.} proposed a master-slave transmission-based concurrent link scheduling algorithm in \cite{wang2021concurrent} to improve the saturation throughput and reduce the average energy consumption. In \cite{fan2020link} and \cite{zhang2021load}, authors focused on optimizing channel utilization and network capacity through link scheduling algorithms. The former assumes that all underwater nodes possess identical transmit power and aims to maximize the number of communications. The latter, on the other hand, categorized transmitters into different and identical directions. Different direction nodes are scheduled using slot-based protocols, while identical direction nodes are scheduled using a reinforcement learning-based competitive protocol with a back-off mechanism.

Power allocation is adopted as a practical means to optimize the overall performance, such as network capacity and energy consumption \cite{islam2022survey}\cite{sun2023bargain}. Researchers are eager to enable more simultaneous communications with limited energy supplies and time-varying interference through power allocation. Jornet \textit{et al.} emphasized in \cite{jornet2010joint} that battery-powered nodes must minimize energy consumption without compromising network connectivity and transmission reliability. \cite{yu2018power} seeks to enhance the energy efficiency of underwater nodes and prolong the network lifetime. Their proposed node power allocation scheme determines the optimal transmit power based on the communication distance and remaining energy. In \cite{zhang2021scalable}, Zhang \textit{et al.} investigated the effect of unfair transmissions on network performance under conditions of limited energy supplies. The authors present a distributed fair power allocation approach to improve spatial reuse and network capacity for resource-constrained underwater networks.

The integration of link scheduling and power allocation is a common approach to concurrently ensure communication quality and reduce energy consumption in unstable multi-user systems. \cite{cruz2003optimal} seeks to find the optimal JSA policy that minimizes the average energy consumption in a wireless network, subject to given constraints on the minimum average data rate per link and peak transmission power per node. ElBatt \textit{et al.} developed a two-phase solution that combines link scheduling and power control to address the multiple access problem in wireless ad-hoc networks \cite{elbatt2004joint}. The proposed algorithm effectively increases single-hop throughput while decreasing energy consumption. \cite{le2014joint} aims to provide effective and eco-friendly communication for underwater cognitive acoustic networks. Their strategy maximizes frequency efficiency while avoiding interfering with both natural and artificial sound systems. Additionally, \cite{wang2022joint} presents a joint approach to increase the throughput of UWSNs by leveraging the slow propagation speed of acoustic signals.


Reinforcement learning is a learning technique in which multiple agents learn how to act by interacting with the environment \cite{gou2023achieving}. Deep RL (DRL) leverages deep neural network as universal function approximators to estimate action values in Q-learning, thereby solving the high dimensionality problem of Q-learning \cite{mnih2015human}. In addressing the challenge of link scheduling without prior environmental knowledge in underwater scenarios, \cite{valerio2015reinforcement} employs reinforcement learning to strike a balance between energy consumption and packet delivery latency. In \cite{wang2019self}, Wang \textit{et al.} introduces a distributed resource allocation algorithm using the Q-learning method to maximize network capacity and maintain communication link quality. In \cite{gou2021achieving} and \cite{zhang2021udarmf}, RL methods were used to optimize the performance of UWSNs. While \cite{gou2021achieving} aims to achieve time-sharing and spatial-reuse in UWSNs with communication fairness, it requires optimization of the delivery ratio. To maximize local capacity and global spatial reuse, \cite{zhang2021udarmf} presents a distributed and adaptive resource management model, termed as UDARMF. However, UDARMF lacks considerations for mobile nodes and communication fairness. Furthermore, existing RL solutions do not adequately address the imperfect nature of UWSNs, and strategies for offline or outdated intelligent nodes are not considered. To accommodate to IC-UWSNs, we propose a joint link scheduling and power allocation scheme based on deep MARL to provide fair, efficient, and reliable communication.

\vspace{-1em}
\section{System Model}\label{sec:sys}

\subsection{Network and Acoustic Channel Model}\label{sec:system}

\begin{figure}[htbp]
\begin{center}
\includegraphics[width=3.2in]{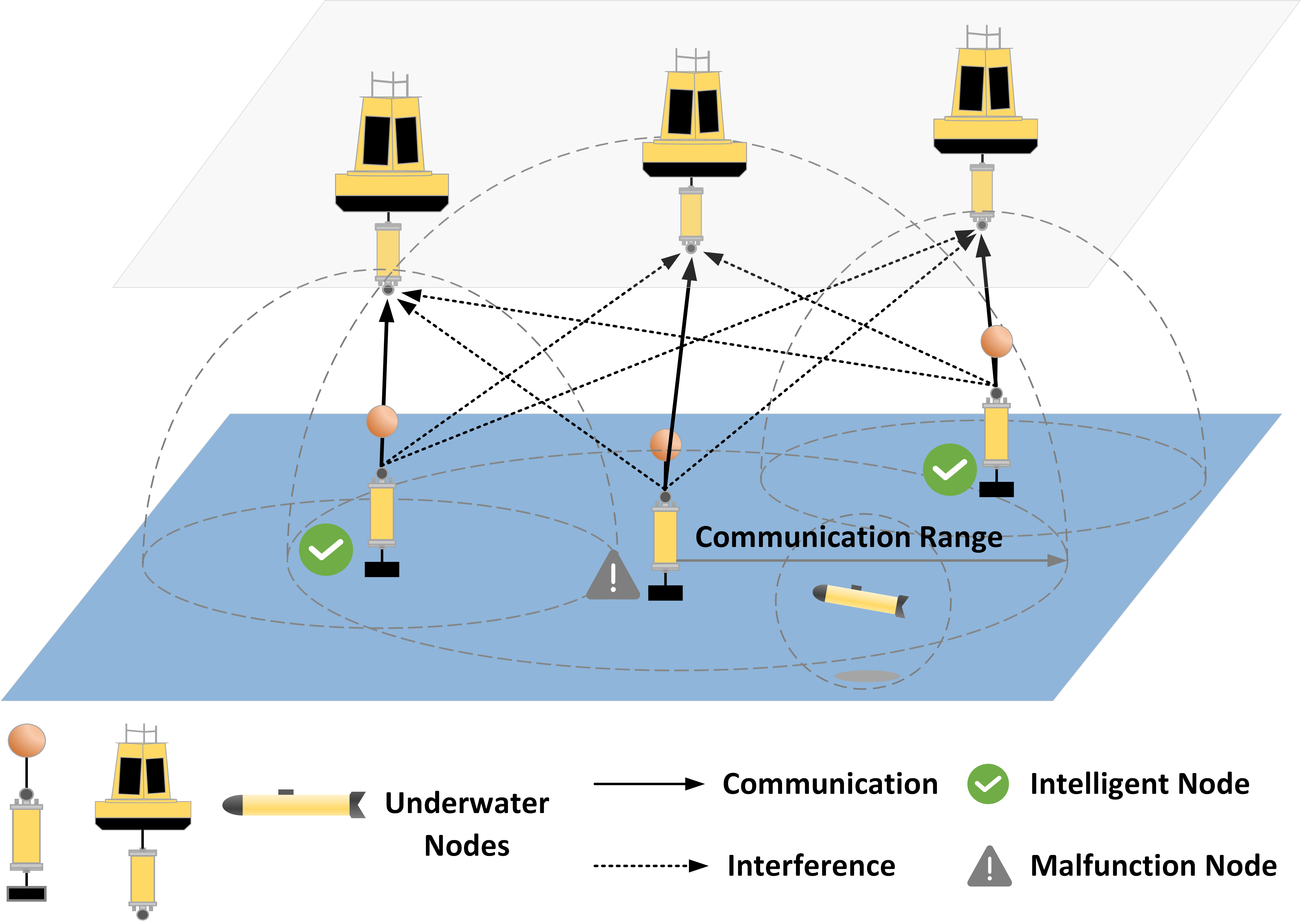}
\vspace{-1em}
\caption{An imperfect and energy-constrained underwater wireless sensor network consisting of multiple transmitter-receiver pairs.}
\label{fig:imUWSN}
\end{center}
\end{figure}

This paper considers the imperfect and energy-constrained UWSNs (IC-UWSNs), as shown in Fig. \ref{fig:imUWSN}. There are $N$ underwater nodes randomly distributed in 3D underwater space. The transmitters are equipped with half-duplex and omnidirectional acoustic modems, and share the same acoustic channel. Let $\mathcal{N}\!\triangleq\!\{n|n\!=\!1,2,\ldots,N\}$ be the set of underwater nodes, $\mathcal{N}_{send}\!\subseteq\!\mathcal{N}$ be the set of transmitters in the considered system, and $\mathcal{N}_{send}^t\!\subseteq\!\mathcal{N}_{send}$ be the set of transmitters scheduled to send at time slot $t$. $n_i\in \mathcal{N}_{send}^t$ denotes the scheduled transmitter, and $\mathcal{N}^t_{i,recv}\!\subseteq\!\mathcal{N}^{-i}$ represents the set of intended receivers. Specifically, for transmitter $n_i$, $\mathcal{N}^t_{i,recv}\!=\!\{n_{\tilde{i}}\}_{i\neq\tilde{i}}$ denotes the unicast communication scenario, $\mathcal{N}^t_{i,recv}\!\subset\!\mathcal{N}^{-i}$ denotes the multicast scenario, and $\mathcal{N}^t_{i,recv}\!=\!\mathcal{N}^{-i}$ describes the broadcast scenario. In each transmission slot, the transmitter exclusively communicates with its designated receiver(s) through the data channel, and the receiver is permitted to send communication feedbacks through the control channel. 

\vspace{-1em}
\vspace{-1em}
In this paper, the underwater nodes that work cooperatively to improve network performance are defined as \textit{intelligent nodes}, which perform learning-based link scheduling and power allocation. The nodes that fail to load the intelligent algorithms due to hardware or software failures are defined as \textit{malfunction nodes}. The malfunction nodes behave irrationally, which may generate suppressive interference to other concurrent communications. The underwater nodes move passively due to water currents at a speed of $c$ in meters per second (m/s), and the mobility model $\mathcal{M}_{k,c}$ is as given in \cite{he2020trust}. $\boldsymbol{X}_{i}^{t+1}\!=\!\mathcal{M}_{k,c}(\boldsymbol{X}_{i}^{t})$, where $\boldsymbol{X}_{i}^{t}\!=\!(x_{i}^{t},y_{i}^{t},z_{i}^{t})$ is $n_i$'s 3D Cartesian coordinates at $t$ slot. $d_{i,\tilde{i}}$ is the distance between $n_i$ and $n_{\tilde{i}}$, and $l_{i,\tilde{i}}$ denotes the communication link between $n_i$ and $n_{\tilde{i}}$ (if existed), which will be defined in Section \ref{sec:cons}. Perfect knowledge of node locations is available for all underwater nodes. Note that the approach proposed in this paper is not restricted to a specific kinematic model. However, the model should at least take into account the impact of water currents on underwater nodes.

The underwater acoustic channel exhibits high attenuation, significant communication interference from neighboring acoustic entities and simultaneous transmissions, and constant ambient noise. Signal attenuation, denoted as $A(d_{i,\tilde{i}},f)$, is influenced by both the communication distance $d_{i,\tilde{i}}$ and the carrier frequency $f$. The Urick propagation model \cite{urick1983principles} characterizes this attenuation on each link by utilizing mathematical expressions for distance-related spreading loss and frequency-related absorption loss, enabling researchers to estimate received signal strength \cite{morozs2020channel}. Within the Urick model, the received signal strength is calculated using geometric spreading loss and Thorp absorption formula when $f$ is in kHz \cite{stojanovic2007relationship}. For lower frequencies that above a few hundred Hz, an approximation for absorption loss can be found in \cite[Ch. 3]{etter2018underwater}.

In this paper, we use the received signal-to-interference-plus-noise ratio (SINR) $\gamma_{i,\tilde{i}}$ to evaluate the signal strength from $n_{i}$ to $n_{\tilde{i}}$ over a distance $d_{i,\tilde{i}}$, as (\ref{equ:SINR}) \cite{stojanovic2007relationship},
\begin{equation}
\gamma_{i,\tilde{i}}=\frac{\eta_{0} p_{i}A_{i,\tilde{i}}(d_{i,\tilde{i}},f)^{-1}}{\eta_{0}\sum_{j \in \mathcal{N}_{send}^{-i}} p_{j}A_{j,\tilde{i}}(d_{j,\tilde{i}},f)^{-1}+I_{s}+I_{a}}
\label{equ:SINR}
\end{equation}
where $\eta_{0}$ represents the transducer efficiency in converting electrical power to acoustic power, $p_{i},p_{j}\!\in\!\mathcal{P}$ are the transmit power of $n_{i}$ and $n_j$. $\mathcal{P}$ denotes the set of available transmit power levels for the underwater nodes. $I_{s}$ refers to the interference caused by surrounding acoustic entities, while $I_{a}$ represents the ambient noise generated by the acoustic channels that is discussed in \cite[Ch. 15]{dhanak2016springer}. Taking into account the interference from surrounding acoustic entities, the ambient noise, and the propagation features of the acoustic channel, it is expected that each transmission will be received successfully by its intended receiver, while causing only gentle interference to other simultaneous transmissions. Transmitting at excessively high power levels may impair other ongoing communications \cite{pompili2009cdma}, while insufficient power levels can lead to local communication failures by violating communication requirements. For the convenience of reference, Table \ref{tab:para} provides the notations that are utilized throughout the paper.
\vspace{-1em}
\begin{table}[htp]
\caption{Notation Description}
\vspace{-1em}
\begin{center}
\begin{tabular}{ll}
\hline
Notation&Descriptions\\
\hline
$\mathcal{N}$, $N$&Node set, number of underwater nodes\\
$\mathcal{P}$, $p$&Power set, transmit power\\
$n_i$,$\tilde{n_i}$, $l_{i,\tilde{i}}$&Transmitter, receiver, communication link\\
$s_{i,\tilde{i}}$, $re_{i,\tilde{i}}$&Behavior indicator\\
$\gamma_{i,\tilde{i}}$, $\gamma^{th}$&Received SINR, communication threshold\\
$\delta^{0}$, $\delta_{dur}$&Required lifetime, slot duration\\
$f, B$&Carrier frequency, bandwidth\\
$c_{i,\tilde{i}}, \rho_{\mathcal{N}}$&Data rate, throughput\\
$e_i$/$E_i$&Energy consumption\\
$\kappa_{i}$, $\epsilon$&Malfunction state, malfunction rate\\
\hline
\end{tabular}
\end{center}
\label{tab:para}
\end{table}
\vspace{-2em}

\subsection{Slot Model}

In Slotted-UWSNs, the time domain of the acoustic channel is partitioned into consecutive transmission slots \cite{song2019optimizing}. At the beginning of each time slot, the scheduled nodes send to their intended receiver(s), while unscheduled nodes await their designated transmission slots. It should be noted that within a single transmission slot, a node cannot serve as both a transmitter and receiver simultaneously due to inherent half-duplex restrictions. Following a similar approach to \cite{song2019optimizing} and \cite{zhu2014toward}, the duration $T_{slot}$ of each time slot is defined in (\ref{equ:slot}), considering the transmission delay $T_{tran}$ and the long propagation delay $T_{prop}$ in the underwater acoustic channel.
\begin{equation}
T_{slot}=T_{tran}+T_{prop}+T_{guard}
\label{equ:slot}
\end{equation}
To prevent inter-slot interference, $T_{prop}$ is defined as the maximum propagation delay among all transmitter-receiver pairs in the network. $T_{tran}$ is determined based on the design of specific applications. In this paper, the propagation delay is determined using the approach outlined in \cite{morozs2020channel}, which utilizes the BELLHOP beam tracing method to account for the time-varying communication environment and non-linear propagation path. Additionally, $T_{guard}$ is introduced to compensate for other potential delays (e.g., processing delay) and any additional time that may be required for various tasks \cite{zhu2014toward}.

\subsection{Node Malfunction Model}

In real-world underwater communication systems, unexpected malfunctions may occur in underwater nodes, leading to a significant degradation in the performance of UWSNs. In the communication scenarios considered in this paper, each underwater node can exist in two possible states: normal and malfunction. Specifically, underwater nodes are either in a normal state or in a malfunction state, and the states of the nodes are independent of each other. Therefore, building upon the system failure model as presented in \cite{zhao2022coach}, this paper assumes that each node follows a Bernoulli distribution with a parameter $\epsilon$, representing the node malfunction rate. The malfunction state of the node is denoted by a binary indicator $\kappa$. When $\kappa_{i}\!=\!0$, it indicates that $n_i$ is an intelligent node aiming to maximize network utility cooperatively through the configured intelligent algorithms. The set of intelligent underwater nodes is represented by $\mathcal{N}_{in}$. On the other hand, $\kappa_{i}\!=\!1$ signifies that $n_i$ is a malfunction node that ceases transmission, disregarding lifetime requirements and network performance. It is assumed that a node in the normal state may experience a malfunction in any time slot, and once a node degrades into the malfunction state, it cannot regain its functionality. This assumption is consistent with the high deployment and recovery costs associated with underwater nodes \cite{wei2016power}.

\subsection{Limitations and Constraints}\label{sec:cons}

Before optimizing network performance through link scheduling and power allocation schemes, the limitations and challenges of underwater applications are discussed. \textit{First}, as underwater nodes are typically battery-powered and difficult to recharge after deployment, network lifetime becomes a major concern in optimizing the performance of UWSNs. Due to the costly and time-consuming nature of underwater platform deployment \cite{wei2016power}, UWSNs often need to provide services for a predetermined lifetime $\delta^{0}$. In the considered IC-UWSNs system, the malfunction nodes behave irrationally, and the lifetime of malfunction nodes cannot be optimized by the method proposed in this paper. Therefore, the focus of this paper is primarily on the resource allocation of intelligent nodes to satisfy their lifetime requirements. Similar to definitions described in \cite{zhang2021udarmf} and \cite{islam2022lifetime}, the network lifetime $\delta_{\mathcal{N}}$ is defined in (\ref{equ:lifetimeDef}) as the time until the first intelligent node exhausts its battery. 
\begin{equation}
\delta_{\mathcal{N}}=\min_{i \in \mathcal{N}_{in}}\{\delta_{i}\}
\label{equ:lifetimeDef}
\end{equation}
where $\delta_{i}$ is the lifetime of $n_i$. Consider $E^{0}$ as the identical battery capacity for all underwater nodes in the IC-UWSNs. Let $e_{i}^{t}\!=\!p_{i}^{t} \times T_{tran}$ represent the energy consumption of $n_i$ at time slot $t$, and $E_{i}^{t}\!=\!\sum_{1}^{t}e_{i}^{t}$ denotes the total energy consumption in the past $t$ time slots, which shall not exceed the total battery capacity.

\textit{Second}, the \textit{Physical Model} \cite{gupta2000capacity} is adopted in this paper to model the successful reception of a transmission, where a minimum SINR $\gamma^{th}$ is required. Taking an information-theoretic viewpoint, if the SINR at the receiver $\gamma_{i,\tilde{i}}$ exceeds the predetermined threshold $\gamma^{th}$, the received signal can be decoded with an acceptable bit error rate, and the interference as well as noise are assumed to be Gaussian \cite{stamatiou2011throughput}. A transmission is successful only if $\gamma_{i,\tilde{i}} \ge \gamma^{th}$ is satisfied. The SINR threshold reflects the sensitivity of the acoustic receiver’s signal detection and decoding capability. This principle is consistent with the actual design of acoustic communication transceivers \cite{freitag2005whoi}. A communication link $l_{i,\tilde{i}}$ between $n_i$ and $n_{\tilde{i}}$ is established only when the signal strength of $n_i$'s transmission experienced at $n_{\tilde{i}}$ is sufficient for decoding. Conversely, if the signal strength is insufficient, no communication link is formed between them.

For each communication link $l_{i,\tilde{i}}$, the send and receive behaviors are defined by $s_{i}$ and $re_{i,\tilde{i}}$, respectively. $s_{i}^{t}$=1 if $n_i$ is scheduled to send at $t$ slot, otherwise it remains \textit{idle}. If $n_{\tilde{i}}$ successfully received from $n_i$ at $t$ slot, $re_{i,\tilde{i}}^{t}$=1; otherwise $re_{i,\tilde{i}}^{t}$=0. Specifically in broadcast scenario, for $\tilde{i}\!\in\!\mathcal{N}^{-i}$, $s_{i,\tilde{i}}^{t}$=1 when $n_i$ is scheduled. Note that there are two possible reasons for \textit{vacant} receiver (i.e., $re_{i,\tilde{i}}^{t}$=0). First, $n_i$ is idle at $t$ slot. The second possibility is that the strength of the received signal is inadequate for decoding due to interference from other concurrent communications. The former decreases spatial reuse, whereas the latter diminishes network reliability and wastes limited energy resources.

Based on the above analysis, the following constraints are crucial considerations in the design of link scheduling and power allocation schemes. Firstly, $\gamma_{i,\tilde{i}}$ should satisfy the communication threshold $\gamma^{th}$, ensuring that the received signal strength is robust enough for successful decoding at the receiver. Secondly, each transmission should avoid causing suppressive interference, leading to inadequate signal strength for decoding in other ongoing communications. Achieving this objective involves judiciously scheduling communication links and allocating transmit power to the scheduled transmitters \cite{elbatt2004joint}.

Depending on the Shannon formula \cite{stojanovic2007relationship}, given a communication link $l_{i,\tilde{i}}$, the achievable data rate $c_{i,\tilde{i}}^{t}$ in bits-per-second (bps) is calculated as (\ref{equ:capacity}).
\begin{equation}
c_{i,\tilde{i}}^{t}=\left\{
\begin{array}{ll}
B\times \log_{2}(1+{\gamma}_{i,\tilde{i}}^{t}), & \gamma_{i,\tilde{i}}^{t} \ge  \gamma^{th}\\
0, & otherwise
\end{array} \right.
\label{equ:capacity}
\end{equation}
where $B$ in Hz is the bandwidth. It is necessary to increase network throughput for better supporting various underwater applications. As described in Section \ref{sec:system}, this paper mainly considers two distinct communication scenarios, i.e., unicast communication scenario and broadcast communication scenario. Applications falling into the former scenario aim to deliver a larger volume of data within a specific period of time, while broadcast-based applications prioritize maximizing successful receptions for enhanced network-wide acknowledgments. Therefore, the definitions of throughput in these scenarios differ slightly. In unicast scenario, throughput is defined as the ratio of the number of bits successfully received at the receivers to the network lifetime $\delta_\mathcal{N}$, as indicated in \cite{diamant2018fair} and expressed in ((\ref{equ:netThroughput_I}).
\begin{equation}
\rho_{\mathcal{N}}=\frac{1}{\delta_{\mathcal{N}}\times T_{slot}}\sum\nolimits_{i \in \mathcal{N}_{send}}\sum\nolimits_{t=1}^{\delta_{\mathcal{N}}}(c_{i,\tilde{i}}^{t} \times T_{tran})
\label{equ:netThroughput_I}
\end{equation}
where $\delta_{tran}$ denotes the duration of each transmission.

In the broadcast scenario, throughput is defined to assess the average number of successful deliveries across all nodes during the network lifetime $\delta_\mathcal{N}$\cite{diamant2013robust}, as given in (\ref{equ:netThroughput_II}). 
\begin{equation}
\rho_\mathcal{N}=\frac{1}{\delta_\mathcal{N}\times T_{slot}}\sum\nolimits_{i\in\mathcal{N}_{send}}\sum\nolimits_{\tilde{i}\in\mathcal{N}^{-i}}\sum\nolimits_{t=1}^{\delta_\mathcal{N}}{re_{i,\tilde{i}}^{t}}
\label{equ:netThroughput_II}
\end{equation}
where $re_{i,\tilde{i}}^{t}\!\in\!\{0,1\}$ signifies whether the broadcast packet from $n_i$ is successfully received by $n_{\tilde{i}}$ at $t$ slot. Specifically, $re_{i,\tilde{i}}^{t}\!=\!1$ indicates a successful reception, $re_{i,\tilde{i}}^{t}\!=\!0$ otherwise.
\vspace{-1em}

\subsection{Problem Formulation}\label{sec:pf}

We aim to provide fair, efficient, and reliable (FER) communications in imperfect and energy-constrained underwater wireless sensor networks (IC-UWSNs). An appropriate number of links in the network are scheduled to send, and the scheduled nodes rationally select their transmit power to cooperatively satisfy the constraints in $\S$\ref{sec:cons}. Given the challenging channel conditions, constrained energy supply, and potential node malfunctions in underwater networks, the maximization of the long-term network utility is of significant importance, particularly from the perspective of network managers. In this paper, the network utility comprises spatial reuse utility, communication fairness utility, and ineffective communication utility.

\textit{1) Spatial reuse utility.} Due to the complexity of the acoustic channel, network throughput and information freshness of UWSNs considerably fall short of expectations when supporting diverse underwater applications. Consequently, network capacity should be maximized to enhance network efficacy. Exploiting the spatial separation of the underwater nodes to maximize the number of concurrent communications in the network is a viable solution. In this paper, \textit{spatial reuse index} is defined to evaluate the level of spatial reuse among underwater nodes sharing the acoustic channel.
\begin{defn}
(Spatial reuse index ($\mathcal{I}_{\mathcal{N}}^{(Spa)}$)). For $i \in \mathcal{N}$, spatial reuse index $\mathcal{I}_{\mathcal{N}}^{(Spa)}: \mathbb{R}_{0,+}^{\sum_{i\in\mathcal{N}_{send}}|\mathcal{N}_{i,recv}^t|} \to \mathbb{R}_{0,+}$ is defined as the average number of successful communications at each time slot, which is given by (\ref{equ:Spa}).
\begin{equation}
\mathcal{I}_{\mathcal{N}}^{(Spa)}(t)=\frac{\sum_{i\in\mathcal{N}_{send}}\sum_{\tilde{i}\in\mathcal{N}_{i,recv}^t}re_{i,\tilde{i}}^{t}}{\sum_{i\in\mathcal{N}_{send}}|\mathcal{N}_{i,recv}^t|}
\label{equ:Spa}
\end{equation}
\label{defn:CI}
\end{defn}
\vspace{-1em}
This definition shows that $\mathcal{I}_{\mathcal{N}}^{(Spa)}\!\in\! [0,1]$ in both scenarios, and a higher $\mathcal{I}_{\mathcal{N}}^{(Spa)}$ indicates that more nodes accessed the shared acoustic channel and made successful communications at $t$ slot. $\mathcal{I}_{\mathcal{N}}^{(Spa)}\!=\!0$ represents the worst case where all receivers are vacant. Conversely, $\mathcal{I}_{\mathcal{N}}^{(Spa)}\!=\!1$ represents the best case where all nodes share the acoustic channel equally. We aim to maximize network throughput $\rho_\mathcal{N}$ over the network lifetime. This can be achieved by scheduling the communication links in the network and allocating appropriate transmit power levels to the scheduled nodes. Thus, the spatial reuse utility $\mathcal{U}_{\mathcal{N}}^{(Spa)}$ can be designed as (\ref{equ:obj1}),
\begin{equation}
\mathcal{U}_{\mathcal{N}}^{(Spa)}=\frac{1}{\delta_{\mathcal{N}}}\sum\nolimits_{t=1}^{\delta_{\mathcal{N}}}\mathcal{I}_{\mathcal{N}}^{(Spa)}(t)\label{equ:obj1}
\end{equation}

\textit{2) Communication fairness utility.} Enhancing spatial reuse is essential for optimizing network performance while ensuring communication fairness among underwater nodes is equally imperative. This statement can be justified for two key reasons. \textit{First}, given the limited energy supplies of underwater nodes, those transmitting frequently are likely to deplete batteries sooner. This not only cut short network lifetime but also reduces network availability. \textit{Second}, the accuracy of information degrades over time. For nodes with infrequent transmission opportunities, the intolerable delivery delay renders the data useless (or even expired in some time-sensitive applications). By considering the UWSNs as multi-user systems, this paper defines the \textit{communication fairness index} with Jain's fairness index \cite{jain2008art}.
\begin{defn}
(Communication fairness index $\mathcal{I}_{\mathcal{N},h}^{(Fair)}$). For $i \in \mathcal{N}$, $\mathcal{I}_{\mathcal{N},h}^{(Fair)}$ is defined as a function of the variety of deliveries among underwater nodes during the past $h$ slots, which is given by (\ref{equ:Fair1}),
\label{defn:lf}
\end{defn}
\vspace{-0.5em}
\begin{equation}
\mathcal{I}_{\mathcal{N},h}^{(Fair)}(t)=\frac{(\sum_{i\in \mathcal{N}_{send}}\sum_{\tilde{i}\in\mathcal{N}_{i,recv}}\sum_{t-h+1}^{t}re_{i,\tilde{i}}^{t})^{2}}{|\mathcal{N}_{send}|\sum_{i\in \mathcal{N}_{send}}(\sum_{\tilde{i}\in\mathcal{N}_{i,recv}}\sum_{t-h+1}^{t}re_{i,\tilde{i}}^{t})^{2}}
\label{equ:Fair1}
\end{equation}
Normally, $\mathcal{I}_{\mathcal{N}}^{(Fair)}$ is continuous and lies in $[1/N,1]$. At each time slot, $\mathcal{I}_{\mathcal{N}}^{(Fair)}\!=\!1$ describes the fairest scenario where all nodes communicate equally and successfully with their intended receivers, and $\mathcal{I}_{\mathcal{N}}^{(Fair)}\!=\!1/N$ refers to the least fair allocation where a single node occupies the acoustic channel. Specifically, in this paper, we define $\mathcal{I}_{\mathcal{N}}^{(Fair)}\!=\!0$ when there are no successful communications in the whole system, which is practical with respect to the complex acoustic channel conditions. We aim to achieve fair communications in IC-UWSNs by maximizing the network communication fairness index. Thus, the communication fairness utility $\mathcal{U}_{\mathcal{N},h}^{(Fair)}$ can be designed as (\ref{equ:obj2}).
\begin{equation}
\mathcal{U}_{\mathcal{N}}^{(Fair)}=\frac{(\sum_{i\in \mathcal{N}_{send}}\sum_{\tilde{i}\in\mathcal{N}_{i,recv}}\sum_{t=1}^{\delta_{\mathcal{N}}}re_{i,\tilde{i}}^{t})^{2}}{|\mathcal{N}_{send}|\sum_{i\in \mathcal{N}_{send}}(\sum_{\tilde{i}\in\mathcal{N}_{i,recv}}\sum_{t=1}^{\delta_{\mathcal{N}}}re_{i,\tilde{i}}^{t})^{2}}
\label{equ:obj2}
\end{equation}

3) \textit{Ineffective communication utility.} Because the underwater nodes are difficult to recharge or replace the battery after deployment, the UWSNs are characterized by constrained energy supplies. In all communication scenarios (i.e., unicast, multicast, and broadcast scenario), when $s_{i}^{t}\!=\!1$ and $re_{i,\tilde{i}}^{t}\!=\!0$, the communication at $l_{i,\tilde{i}}^{t}$ is described as \textit{ineffective communications}, which not only wastes energy but also causes interference to other concurrent communications in the network. The \textit{ineffective communication index} is defined as follows.
\begin{defn}
(Ineffective communication index ($\mathcal{I}_{\mathcal{N}}^{(Ief)}$)). For $i \in \mathcal{N}_{send}^t$, ineffective communication index $\mathcal{I}_{\mathcal{N}}^{(Ief)}: \mathbb{R}_{0,+}^{\sum_{i\in\mathcal{N}_{send}^t}|\mathcal{N}_{i,recv}^t|} \!\to\! \mathbb{R}_{0,+}$ is defined as the ratio of ineffective communication to the overall number of transmissions at time slot $t$, as given by (\ref{equ:Ief}).
\begin{equation}
\mathcal{I}_{\mathcal{N}}^{(Ief)}(t)=\frac{\sum_{i\in\mathcal{N}_{send}^t}\sum_{\tilde{i}\in\mathcal{N}_{i,recv}^t}(re_{i,\tilde{i}}^{t}-s_{i,\tilde{i}}^{t})}{\sum_{i\in\mathcal{N}_{send}^t}\sum_{\tilde{i}\in\mathcal{N}_{i,recv}^t}s_{i,\tilde{i}}^{t}}
\label{equ:Ief}
\end{equation}
\label{defn:Ief}
\end{defn}
\vspace{-1em}
This definition indicates that $\mathcal{I}_{\mathcal{N}}^{(Ief)}$ lies in $[-1,0]$ and is related to all scheduled communication links in the network. $\mathcal{I}_{\mathcal{N}}^{(Ief)}(t)\!=\!-1$ if all transmissions fail at time slot $t$, and $\mathcal{I}_{\mathcal{N}}^{(Ief)}(t)\!=\!0$ refers to the best situation where all transmissions from the scheduled nodes are successfully received at the intended receiver(s). To efficiently utilize the limited energy of the underwater nodes, the average number of ineffective communications during the network lifetime should be minimized. Thus, the ineffective communication utility $\mathcal{U}_{\mathcal{N}}^{(Ief)}$ can be designed as (\ref{equ:obj3}).
\begin{equation}
\mathcal{U}_{\mathcal{N}}^{(Ief)}=\frac{1}{\delta_{\mathcal{N}}}\sum\nolimits_{t=1}^{\delta_{\mathcal{N}}}\mathcal{I}_{\mathcal{N}}^{(Ief)}(t)
\label{equ:obj3}
\end{equation}

Combining (\ref{equ:obj1}), (\ref{equ:obj2}), and (\ref{equ:obj3}), the long-term network utility $\mathcal{U}_{\mathcal{N}}$ can be expressed as (\ref{equ:netU}),
\begin{equation}
\mathcal{U}_{\mathcal{N}}=\alpha\mathcal{U}_{\mathcal{N}}^{(Spa)}+\beta\mathcal{U}_{\mathcal{N}}^{(Fair)}+\mu\mathcal{U}_{\mathcal{N}}^{(Ief)}
\label{equ:netU}
\end{equation}
where, $\mathcal{U}_{\mathcal{N}}^{(Spa)}$ and $\mathcal{U}_{\mathcal{N}}^{(Fair)}$ range within $[0,1]$, $\mathcal{U}_{\mathcal{N}}^{(Ief)}$ ranges from $[-1,0]$. The weight parameters $\alpha$, $\beta$, and $\mu$ determine the relative importance assigned to each of these detailed network utilities, which are carefully selected based on application needs and communication scenarios. With the system model and underwater constraints presented in $\S$\ref{sec:system}-\ref{sec:cons}, the FERCOP to maximize the long-term network utility is formulated as (\ref{equ:objmax}),
\begin{alignat}{2}
\max_{\mathcal{L},\mathcal{P}}\quad& \mathcal{U}_{\mathcal{N}}\label{equ:objmax}\\
\mbox{s.t.}\quad
& E_{i}^{t} \le E^{0}, \forall i\in \mathcal{N}, \forall t\in [0,\delta_{\mathcal{N}}]\label{equ:conEnergy}\\
& \delta_{\mathcal{N}} \ge \delta^{0}\label{equ:conLife}\\
& \gamma_{i,\tilde{i}}^{t} \ge \gamma^{th}, \forall i\in \mathcal{N}_{send}, \forall t\in [0,\delta_{\mathcal{N}}]\label{equ:conSINR}\\
& p_{i}^{t} \in \mathcal{P}, \forall i\in \mathcal{N}, \forall t\in [0,\delta_{\mathcal{N}}]\label{equ:conPower}\\
& 0 \le \sum\nolimits_{i \in \mathcal{N}}\kappa_{i} \le N \label{equ:conMal}\\
& 0 \le \mathcal{I}_{\mathcal{N}}^{(Spa)}, \mathcal{I}_{\mathcal{N}}^{(Fair)}, -\mathcal{I}_{\mathcal{N}}^{(Ief)} \le 1\label{equ:conFair}
\end{alignat}
where $\mathcal{L} \!\subseteq\! \mathbb{R}^{N \times \delta_{\mathcal{N}}}$ is the matrix of link scheduling throughout the network lifetime. $\mathbf{s}^{t}\!=\!\{s_{i}^{t} | \forall i \!\in\! \mathcal{N}\} \!\in\! \mathcal{L} \!\subseteq\! \mathbb{R}^{N}$ is the set of link scheduling at time slot $t$. $\mathcal{P}\!=\![\boldsymbol{p}_{i}^{\top}]_{i=1}^{N} \!\subseteq\! \mathbb{R}_{0,+}^{N \times \delta_{\mathcal{N}}}$ is the matrix of transmit powers performed by the underwater nodes at each time slot. Constraint (\ref{equ:conEnergy}) bounds that for each underwater node, the total energy consumption should not exceed the battery capacity. Constraint (\ref{equ:conLife}) ensures the the satisfaction of lifetime requirements. Constraint (\ref{equ:conSINR}) gives the minimum signal strength requirement for successful receptions at the receiver. Constraint (\ref{equ:conPower}) ensures that the performed transmit power is available considering the used acoustic modem. And the number of successful communications should not exceed the number of concurrent transmissions. Constraint (\ref{equ:conMal}) gives the limit of malfunction nodes. The number of malfunction nodes is non-negative and will not exceed the number of nodes in the network. Constraint (\ref{equ:conFair}) limits the upper bound of spatial reuse index, communication fairness index, and ineffective communication index. The optimal link scheduling and transmit power allocation scheme $(\mathcal{L},\mathcal{P})^{*}$ is decided as (\ref{equ:MAXutility}).
\begin{equation}
(\mathcal{L},\mathcal{P})^{*}=\arg\max_{(\mathcal{L},\mathcal{P})}\mathcal{U}_{\mathcal{N}}
\label{equ:MAXutility}
\end{equation}

As we evaluate communication fairness among underwater nodes, we focus on scenarios where transmitters operate under a full buffer traffic model. Additionally, all transmitters are homogeneous, possessing identical software/hardware configurations and benefit from comparable channel conditions. No transmitter is disadvantaged due to its transmit power, battery capacity, or deployment. The concepts of shadow zone and convergence zone are beyond the scope of our research.

\section{Design of the ICRL-JSA Scheme}\label{sec:method}
\begin{figure*}[htbp]
\centerline{\includegraphics[width=6.5in]{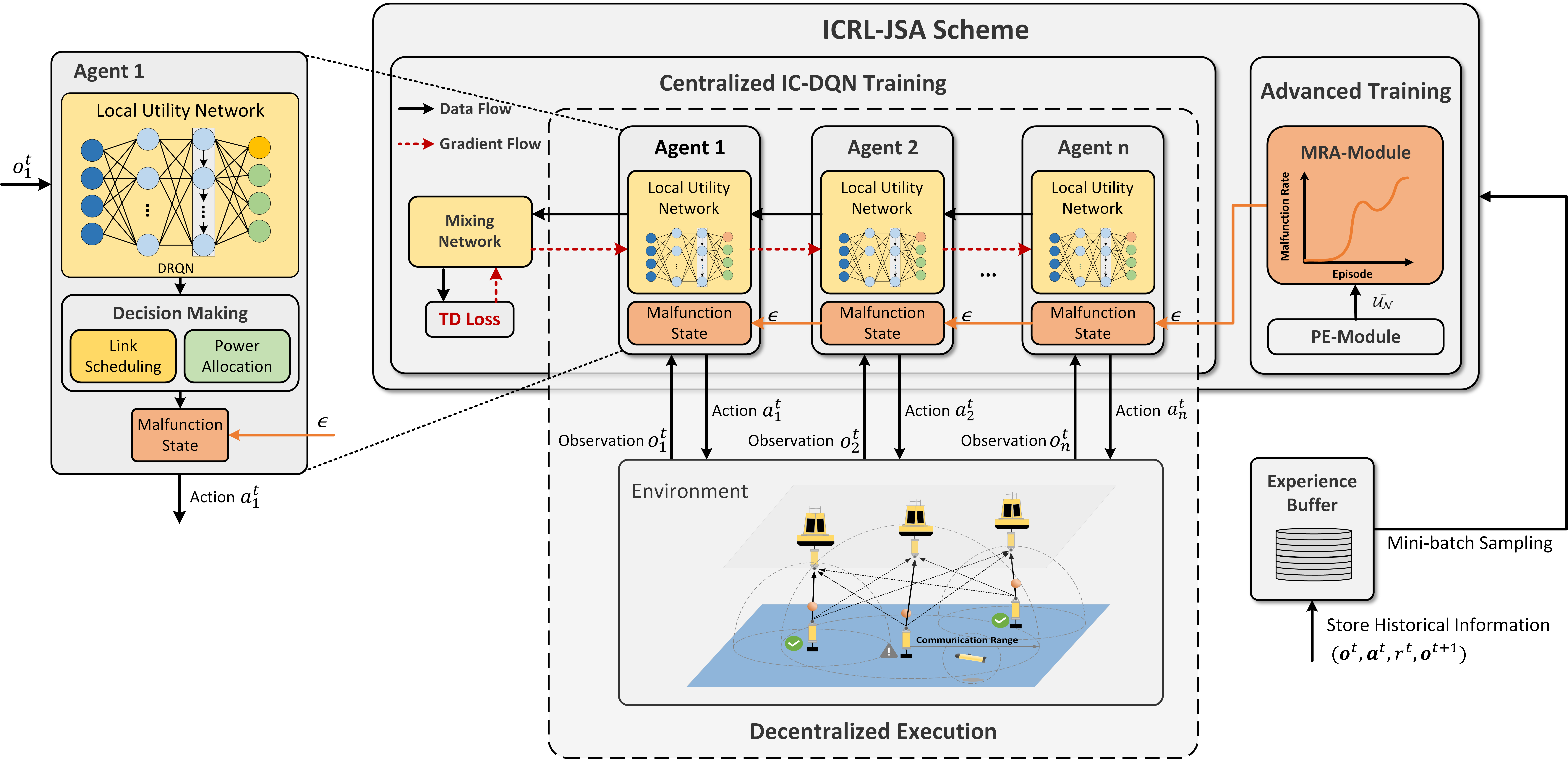}}
\vspace{-1em}
\caption{Working diagram of the proposed ICRL-JSA scheme.}
\label{fig:main}
\end{figure*}

\subsection{Overview of ICRL-JSA}

This section provides an overview of ICRL-JSA, $\S$\ref{sec:DQN} briefly revisits fundamental RL principles, while $\S$\ref{sec:JSA} details the integration of the DQN structure within IC-UWSNs to address constraints associated with energy supply and unexpected node malfunctions. Furthermore, $\S$\ref{sec:ATM} introduces an advanced training mechanism designed to enhance the robustness of agents in ICRL-JSA. ICRL-JSA aims to solve the formulated FERCOP and provide FER-communications for IC-UWSNs by determining (1) \textit{how many} transmitters can transmit in the current communication environment; (2) \textit{which} transmitter to send to improve communication fairness; and (3) \textit{how} the scheduled nodes will transmit to satisfy the communication threshold and avoid suppressing other concurrent transmissions. The first two decisions are related to link scheduling, while the third decision pertains to power allocation. Notably, ICRL-JSA implicitly achieves link scheduling by setting the transmit power of one or more underwater nodes to zero according to the communication environment. Figure \ref{fig:main} illustrates the operation of ICRL-JSA.

In this paper, the centralized training with decentralized execution (CTDE) paradigm \cite{foerster2018counterfactual} is used to train ICRL-JSA agents, considering the partial observability of underwater nodes. In CTDE, the training phase takes place within a simulator with additional state information, while agents must rely on local action-observation histories during execution. The decision to employ the CTDE paradigm over a fully centralized approach is motivated by several key considerations. First, underwater acoustic communication is characterized by limited bandwidth, long propagation delay, and constrained energy supplies. Frequent exchanges of observations and transmission policies between nodes and the controller significantly hinder communication efficiency and deplete already limited bandwidth and energy resources. Secondly, for a centralized controller capable of gathering information from all underwater nodes and making global-wise transmission decisions, scheduling these information presents a formidable challenge. Moreover, underwater acoustic communication exhibits pronounced instability (flickering) \cite{casari2020asuna}. In such scenarios, the central controller encounters difficulty in distinguishing between node malfunctions and communication failures based on received information, further contributing to imperfect and incomplete decision information. The CTDE paradigm, which mitigates the mismatch between local observations and global states without introducing additional communication latency into the system, proves to be well-suited for efficient policy learning in UWSNs.

ICRL-JSA agents employ a deep neural network (i.e., local utility network) to make decisions, including which link to schedule and how to allocate transmit power for assigned transmitters. Specifically, an advanced training mechanism for evaluating model performance (PE-Module, $\S$\ref{sec:PE}) as well as providing an imperfect environment for model training (MRA-Module, $\S$\ref{sec:MRA}) is included in the training phase. The local utility network is trained through trial-and-error simulations with the environment provided by MRA-Module. In the decentralized execution phase, the mixing network, PE-Module, and MRA-Module are removed. The ICRL-JSA agent utilizes its own copy of the learned network, evolves its own hidden state, and selects its actions based on local observations.

ICRL-JSA's design tackles the three main challenges:

(1) \textbf{Constrained energy supplies.} The optimizer must account for the battery capacity of underwater nodes when making scheduling and allocation decisions. Nodes that transmit frequently or at excessive power levels risk depleting their energy prematurely, thereby reducing the network lifetime. However, sparse transmission limits network capacity and result in increased delivery delay. Furthermore, insufficient transmit power may lead to failed transmissions, wasting energy and reducing network reliability. Balancing the optimization of network performance with the fulfillment of network lifetime requirements, especially with limited energy resources, poses a significant challenge.

(2) \textbf{Imperfect UWSNs with complex channel conditions and node malfunctions.} The optimizer must be capable of handling unpredictable communication scenarios and collaborators that may be susceptible to malfunctions. This requires the optimizer to possess the ability to observe the environment and adapt its strategies accordingly.

(3) \textbf{Partial observability.} Owing to the constrained sensing and communication range of underwater nodes, they can only access local partial observations rather than global states. The disparity between local observations and global states introduces inaccuracies in decision-making based on local information.

\subsection{Preliminaries of ICRL-JSA}\label{sec:DQN}

In this paper, a DQN-based agent architecture is used to solve the formulated FERCOP in IC-UWSNs through joint link scheduling and power allocation (JSA). The IC-UWSNs are considered as multi-agent systems in which the distributed agents cooperate to optimize the network performance. The JSA task in IC-UWSNs can be modeled as a decentralized partially observed Markov decision process (Dec-POMDP) defined by a tuple $G\!=\!\langle \mathcal{S}, \mathcal{A}, P, \mathcal{R}, O, \mathcal{O}, N, \gamma \rangle$. $N$ is the number of agents in the system, where the actions are selected and executed simultaneously. $s \!\in\! \mathcal{S}$ is the environment state, $\mathcal{A}$ is the action space. $\mathcal{R}$ is the reward function. Since the agent is partially observable, it can only make local observations $o_{i} \!\in\! \mathcal{O}$, following the observation function $O(s,i)$. At each time slot, the agent $n_{i}$ selects an action $a_{i} \!\sim\! \pi(\cdot|o_{i})$ from $\mathcal{A}$, forming a joint action $\boldsymbol{a}\!=\![\boldsymbol{a_{i}}]_{i=1}^{N} \!\in\! \mathbb{R}^{N}$. After executing $\boldsymbol{a}$, the environment transits to the next state $s^{'}$ according to the transition function $P(s^{'}|s,\boldsymbol{a})$ and all agents receive a team reward $r\!\sim\!\mathcal{R}(s,\boldsymbol{a})$. Given a behavior policy $\pi$ and a state-action pair $(s,a)$, the state-action value is given by $Q_{\pi}(s,a)\!=\!\mathbb{E}[R^{t}|s^{t}\!=\!s,a^{t}\!=\!a]$. The agent aims to find the optimal policy $\pi^{*}\!=\!\arg\max_{a}Q_{\pi}(s,a)$ to maximize the cumulative future reward $R^{t}\!=\!\sum_{t=1}^{T}\gamma^{t-1}r^{t}$, which is discounted by $\gamma\!\in\![0,1]$ at each time slot.

Deep Q-Network (DQN) is one of the commonly used framework in DRL \cite{mnih2015human}, which modifies standard Q-learning to make it suitable for training large neural networks without diverging. First, DQN leverages a deep neural network $\theta$ to parameterize the action-value function in Q-learning and to approximate the optimal action-value. Second, it uses the experience replay mechanism to eliminate correlations in the observation sequence across changes in the distribution. The agents store the experiences in a replay buffer at each time slot, and randomly select a mini-batch of transitions from the buffer to update $\theta$ after each episode. Moreover, DQN adopts two neural networks (i.e.,  evaluation network $\theta$ and target network $\theta^{-}$) to stabilize training. Specifically, the agent evaluates $Q(s,a)$ with $\theta$, and computes a target Q-value with $\theta^{-}$. The evaluation network outputs Q-value for all possible actions, and the action is selected according to certain behavior policy.

In partially observed systems, the policy typically augments the global state $s$ with historical observations $\boldsymbol{\tau}^{t}$. The inclusion of recurrent neural networks in the DQN framework enables the agent to more accurately estimate the environment, thus reducing the disparity between $Q(\boldsymbol{\tau},a;\theta)$ and $Q(s,a;\theta^{-})$, where $\boldsymbol{\tau}_{i}^{t}\!=\!(o_{i}^{1},a_{i}^{1},\ldots,o_{i}^{t},a_{i}^{t})$ represents the sequence of observations and actions over time. DRQN, the combination of recurrent neural network and DQN, is able to handle scenarios with partial observability \cite{hausknecht2015deep}. In this paper, we employ a gated recurrent unit (GRU) denoted to parametrically estimate $\boldsymbol{\tau}^{t}$ \cite{liu2018td}. This adaptation facilitates a more accurate estimation of potential environmental states based on historical observational data, thus diminishing the discrepancy between local observations and the global state.

In the underwater environment, unexpected node malfunctions represent a significant threat to the effectiveness and reliability of UWSNs. Hence, the JSA scheme for IC-UWSNs must exhibit robustness in the face of node malfunctions. The conventional DQN training mechanism typically assumes that RL agents can function normally at all times. However, this assumption is often impractical, particularly in IC-UWSNs. Some underwater nodes may take irrational actions due to unexpected malfunctions, and the other intelligent nodes, initially trained in ideal UWSNs, may struggle to maintain effective cooperation, resulting in a notable degradation in network performance. 

To integrated DQN in IC-UWSNs, termed as IC-DQN, we made three major changes. \textit{First}, during the training phase, we constrained node behavior based on the remaining energy and malfunction state. When a node's remaining energy drops below the energy threshold, it ceases transmission, ultimately leading to network service termination. This satisfies the energy constraint of FERCOP. If a node malfunctions, the intelligent algorithm within it is unable to provide link scheduling and power allocation solutions, forcing the node to stop transmission.

\textit{Second}, to enhance the effectiveness and robustness of the RL model, we introduced an advanced training mechanism ($\S$\ref{sec:ATM}). This mechanism consists of a performance evaluation module (PE-module, $\S$\ref{sec:PE}) and a node malfunction rate adjustment module (MRA-module, $\S$\ref{sec:MRA}). The PE-module formulates performance expectations before training and periodically compares the anticipated utility with the actual performance during the training phase. The MRA-module fine-tunes the node malfunction rate in response to these comparisons, creating a constructive training environment that continuously encourages improvement.

\textit{Third}, we select the optimal model by concurrently assessing the training malfunction rate and the achieved mean episode rewards, in contrast to many RL algorithms that consider only the latter. As the ICRL-JSA scheme adjusts the training malfunction rate based on model performance, models trained with a higher malfunction rate become more adept at handling complex environments.

\subsection{ICRL-JSA Scheme}\label{sec:JSA}

\begin{figure}[htbp]
\begin{center}
\includegraphics[width=2.8in]{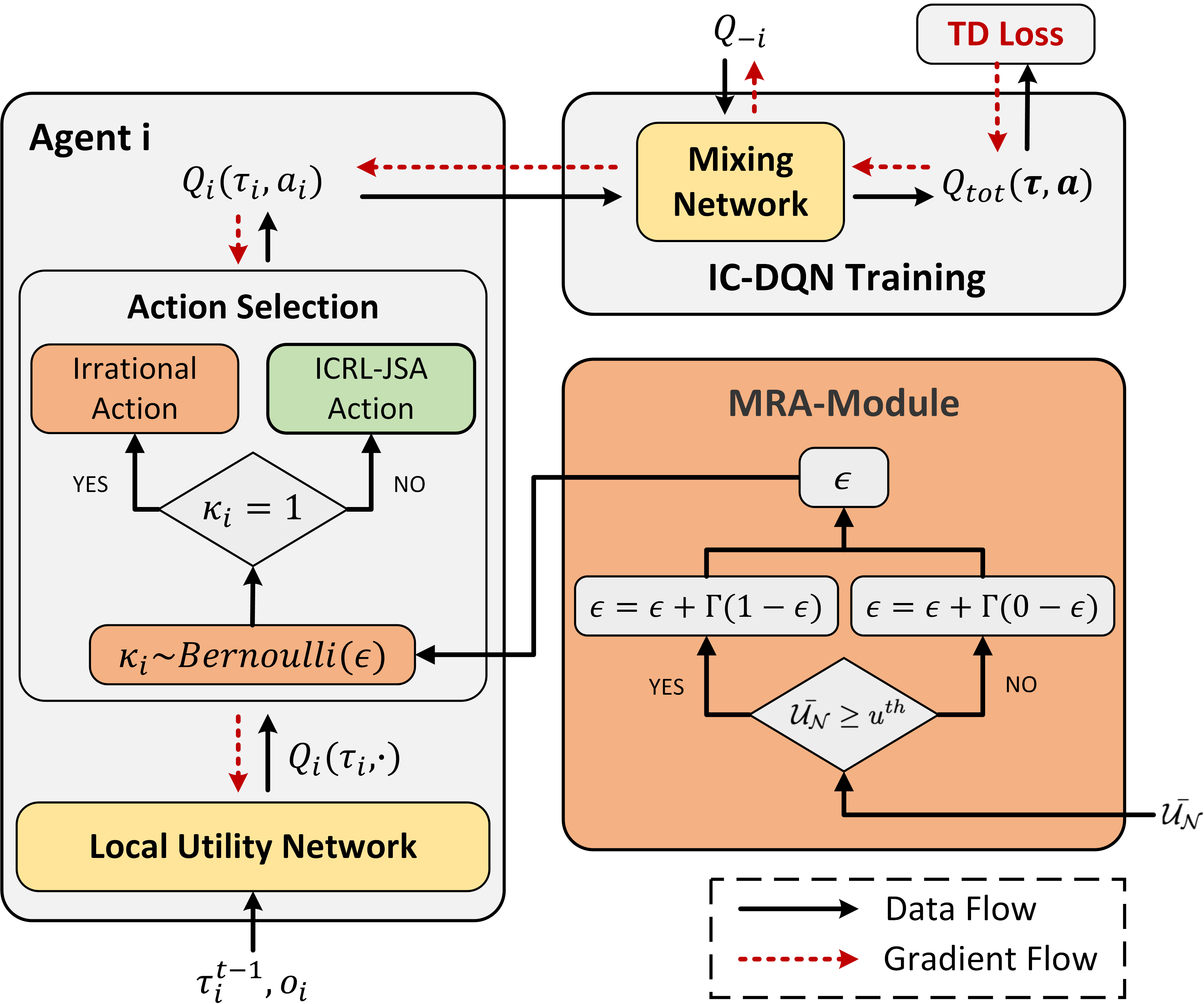}
\vspace{-1em}
\caption{Training procedure of ICRL-JSA.} 
\label{fig:wdJSA}
\end{center}
\end{figure}

We now developed our ICRL-JSA scheme by incorporating the IC-DQN algorithm and the advanced training mechanism. The training procedure for ICRL-JSA is outlined in Fig. \ref{fig:wdJSA} and detailed in Algorithm \ref{alg:modelTraining}. Before applying deep MARL algorithms to solve the formulated FERCOP, we must establish the observation set $\mathcal{O}$, action space $\mathcal{A}$, and reward function $\mathcal{R}$ specialized for IC-UWSNs.
\vspace{-1em}
\vspace{-1em}

\textbf{Environment observations $\mathcal{O}$.} $\mathcal{O}$ is the set of partial observations made by the agents, which are used to make the individual JSA decisions. At time slot $t$, $n_{i}$'s local observations consist of: (1) self-information $\boldsymbol{o}_{i}^{S,t}$, including current position $d_{i}^{t}$, residual energy indicator $e_{i}^{re,t}$ that represents the ratio of currently available energy to the initial battery capacity, last transmission indicator $re_{i,\tilde{i}}^{t-1}$ that indicates whether $n_i$ was scheduled to send in the previous time slot, previous transmission ratio $r_{i,\tilde{i}}^{to,t}$ that evaluates the ratio of times $n_i$ was scheduled during the past $t\!-\!1$ slots, previous transmit power $a_{i}^{t-1}$, and its identifier $i$; (2) position of the intended receiver(s) $\boldsymbol{o}_{i}^{D,t}$; and (3) positions of other transmitters and surrounding acoustic entities $\boldsymbol{o}_{i}^{I,t}$. In particular, $a_{i}^{t-}$ and $i$ are directly obtained from the node. $d_{i}^{t}$, $\boldsymbol{o}_{i}^{D,t}$, and $\boldsymbol{o}_{i}^{I,t}$ are assumed to be available at node $n_i$ throughout the network lifetime, which either pre-configured offline during deployment or acknowledged online through control signals. Future research will focus on addressing the potential degradation in network performance resulting from incomplete or imperfect observations.

\textbf{Action Space $\mathcal{A}$.} $\mathcal{A}$ is mainly constructed by the node transmit power set $\mathcal{P}$. Specifically, link scheduling is implicitly achieved by adjusting the transmit power of one or more underwater nodes to zero at time slot $t$ \cite{elbatt2004joint}, i.e., $p_{i}^{t} \!\ne\! 0$ when $n_{i}$ is scheduled at time slot $t$, $p_{i}^{t}\!=\!0$ otherwise. As a result, $\mathcal{A}\!=\!\mathcal{P}\!\cup\!\{0\}$ is the action space. Since the nodes considered in this paper are identical, $\mathcal{A}\!=\!\mathcal{A}_{i}, i \!\in\! \mathcal{N}$. For an ICRL-JSA node, the executed action is the one that leads to the largest reward based on its partial observations, $a_{i}^{t}\!=\!\arg\max Q(\boldsymbol{o}_{i}^{t}, \boldsymbol{a})$. The malfunction nodes take random actions from $\mathcal{A}$. $\boldsymbol{a}^{t}\!=\![\boldsymbol{a_{i}}^{t}]_{i=1}^{N} \!\in\! \mathbb{R}^{N}$ is the joint action.

\textbf{Reward Function $\mathcal{R}$.} The reward function evaluates the performed actions and provides guidance for future decisions. In the CTDE paradigm, a team reward $r^{t}$ that based on the high-level objective, i.e., providing FER-communications for IC-UWSNs, is given to all agents at each time slot. The ICRL-JSA agents are oriented towards maximizing spatial reuse, ensuring communication fairness, and minimizing the number of failed transmissions, as defined in (\ref{equ:objmax}). To align with network utility, the team reward $r^{t}$ is determined by (\ref{equ:EFRAH}).
\begin{equation}
r^{t} =\left\{
\begin{array}{lcl}
\alpha\mathcal{I}_{\mathcal{N}}^{(Spa)}(t)+\beta\mathcal{I}_{\mathcal{N},h}^{(Fair)}(t)-\mu\varsigma, & & \delta_{\mathcal{N}} \ge \delta^{0} \\
-100, & & otherwise\\
\end{array} \right.
\label{equ:EFRAH}
\end{equation} 
where $h$ is the horizon for assessing communication fairness, $\mathcal{I}_{\mathcal{N}}^{(Spa)}$ and $\mathcal{I}_{\mathcal{N},h}^{(Fair)}$ are as defined in (\ref{equ:obj1}) and (\ref{equ:Fair1}). $\varsigma$ that defined in (\ref{equ:Ief}) serves as the penalty for failed transmissions, which provides guarantees for reliable communication.  In case lifetime requirements are not met, each agent receives a penalty of -100 and the episode is terminated. These penalties are designed to satisfy the network lifetime constraints and enhance the delivery ratio. The adaptability of $h$ is contingent on network deployments, termed as \textit{AH}. Notably, the weight parameters $\alpha$, $\beta$, $\mu$ maintain consistent values with those in (\ref{equ:netU}) and can be adjusted based on the preferences of the network operator and the specific requirements of the application.
\begin{algorithm}
\caption{Training procedure for ICRL-JSA}\label{alg:modelTraining}
\textbf{Initialize:} Experience buffer $\mathcal{B}$, evaluation network $\theta$, target network $\theta^{-}$, update cycle $C$, reward $r$.\\
\For{$episode \gets 1,2,\ldots,M$}{
Initialize $\epsilon$ by Algorithm \ref{alg:ATM}.\\
\For{$t \gets 1, ..., \delta^{0}$}{
\For{$i \gets 1, ..., N$}{
\If {$E_{i}^{t}\le E^{0}$}{
\If {$n_{i}$ is intelligent (i.e., $\kappa_{i}=0$)}{
Select $a_{i}^{t}\in \mathcal{A} \sim \pi_{\theta}$.\label{alg:policyLine}\\
}
\Else{
$n_{i}$ ceases transmission, $a_{i}^{t}=0$.\\
}
}
\Else{
$\delta \gets t$, network service terminated.\\
Finish current episode.\\}
}
Execute joint action $\boldsymbol{a}^{t}$, receive reward $r^{t}$ and new observation $\boldsymbol{o}^{t+1}$.\\
Create and add transition to $\mathcal{B}$.\\
}
Randomly select $b$ transitions from $\mathcal{B}$.\\
\For{$j \gets 1,2,...,b$}
{
Set $y_{j}^{tot}\!\gets\!r_{j}$ if episode terminates at $j\!+\!1$ slot\\
otherwise $y_{j}^{tot}\!=\!r_{j}\!+\!\gamma \max{Q(\boldsymbol{\tau^{t\!+\!1}}, \boldsymbol{a^{t\!+\!1}};\theta^{-})}$.\\
Update $\theta$ by performing a gradient descent step on $Loss(\theta)$ in (\ref{equ:td}).\\
}
\If {current episode is the evaluation episode}{
Update $\theta^{-} \gets \theta_{episode}$
}
Finish current episode.\\
}
$epi^{max}=\arg\max \epsilon_{episode}$.\\
\Return $\theta \gets \arg\max_{episode \ge epi^{max}} r(\theta_{episode})$.
\end{algorithm}

In the centralized training phase, we conceptualize a central controller responsible for aggregating the actions of each agent and distributing the global reward and model parameters among them. The exchange of local observations and parameters is confined to the training phase. Once the training is complete, the agents make decisions independently, using their own copies of the pre-trained model, which does not result in any additional communication resource consumption \cite{xia2021multi}. Training data is generated from agent interactions with the simulated environment. The training procedure of ICRL-JSA is outlined in Algorithm \ref{alg:modelTraining}.

In each transmission slot, $n_i$ makes local observations $o_i^t$ and selects transmit power $a_i^t$ based on the behavior policy $\pi_{\theta}$. Typically, the behavior distribution is often determined by $\varepsilon$-greedy policy \cite{mnih2015human}, which follows the greedy policy with a probability of $1\!-\!\varepsilon$ and selects a random action with a probability of $\varepsilon$\cite{sunehag2018value}. Let $\boldsymbol{a}^{t}\!=\!\{a_{i}^{t}|i\!\in\! \mathcal{N}\}$ be the joint action of all agents at $t$ slot, and $\boldsymbol{\tau}^{t}\!=\!\{\boldsymbol{\tau}_{i}^{t}|i\!\in\! \mathcal{N}\}$ be the joint observations. The virtualized central controller gathers $\boldsymbol{a}^{t}$, $\boldsymbol{o}^{t}$, and $\boldsymbol{o}^{t+1}$ to calculate the team reward $r^{t}$ based on the specified reward function. In each time slot within an episode, the transition of the joint actions $e^{t}\!=\!\{\boldsymbol{o}^{t}, \boldsymbol{a}^{t}, r^{t}, \boldsymbol{o}^{t+1}\}$ is stored in the experience buffer $\mathcal{B}$ that has a finite buffer size $B_r$. The contents of $\mathcal{B}$ are continually overwritten with recent transitions, and for updating parameters $\theta$, a random mini-batch of size $b$ is sampled from $\mathcal{B}$ through Q-learning updates. The central controller holds a mixing network denoted as $f(\cdot)$, which aggregates the outputs of all MARL agents (i.e., Q-value of the selected action) and makes a mixture of them. With the approximated joint action-value $Q_{tot}(\boldsymbol{\tau},\boldsymbol{a})\!=\!f(Q_{1}(\tau_{1}, a_{1}),...,Q_{n}(\tau_{n}, a_{n}))$, a gradient decent step on $(y_{j}^{tot}\!-\!Q_{tot}(\boldsymbol{\tau^{t}}, \boldsymbol{a^t}; \theta))^2$ with respect to the network parameters $\theta$ is performed, as specified in (\ref{equ:td})\cite{mnih2015human}.
\begin{equation}
Loss(\theta)=\sum\nolimits_{j}^{b}(y_{j}^{tot}-Q_{tot}(\boldsymbol{\tau^{t}}, \boldsymbol{a^t}; \theta))^2
\label{equ:td}
\end{equation}
where $y_{j}^{tot}$ represents the target Q-value generated from target network $\theta^{-}$. In this paper, the mixing network $f(\cdot)$ is implemented by value decomposition network (VDN). VDN decomposes the team value function into agent-wise value functions while satisfying the individual-global maximization (IGM) constraint as (\ref{equ:IGM}), by summing up individual value functions $Q_{i}(\tau_{i}, a_{i})$ for each agent.
\begin{equation}
\arg\max_{\boldsymbol{a}}Q_{tot}(\boldsymbol{\tau},\boldsymbol{a})=
\begin{pmatrix}
\arg\max_{a_{1}}Q_{1}(\tau_{1}, a_{1})\\
...\\
\arg\max_{a_{n}}Q_{n}(\tau_{n}, a_{n})
\end{pmatrix}
\label{equ:IGM}
\end{equation} 
Notably, VDN maintains the same number of parameters as independent learning because the summation structure of $f(\cdot)$ requires no additional parameters. Furthermore, while VDN necessitates some centralization during the training phase, it's important to highlight that the trained agents can be deployed independently. This is because each agent, acting in a greedy manner with respect to its local value $Q_{i}(\tau_{i}, a_{i})$, is equivalent to a central controller selecting joint actions by maximizing $Q_{tot}(\boldsymbol{\tau},\boldsymbol{a})$\cite{sunehag2018value}. For the training procedure to be stable and to avoid divergence of the policy, the target network $\theta^{-}$ is updated every $C$ episode by $\theta^{-} \gets \theta$, while remaining unchanged between individual updates.

During the decentralized execution phase, as illustrated in Fig. \ref{fig:main}, the mixing network and the advanced training modules are removed. Each ICRL-JSA agent independently makes decisions based on its local observations with an appropriately trained evaluation network $\theta$, without the need for global information or negotiations with other nodes. Regarding the execution delay, prior research as discussed in \cite{zhao2022deep} has demonstrated that agents can make real-time, distributed decisions using the trained RL model. Moreover, Ye et al. further substantiated in \cite{ye2021scalable} that the average execution time for deep MARL agents falls within the millisecond range, underscoring the effective applicability of the method described in this paper to real-time coordination applications. Hence, the execution delay can be safely disregarded.

\vspace{-1em}
\subsection{Advanced Training Mechanism}\label{sec:ATM}

For an optimizer to effectively generalize in IC-UWSNs, the training episodes must incorporate a variety of malfunction rate levels. To this end, we have proposed an advanced training mechanism. Instead of training customized models for each possible scenario, ICRL-JSA with the advanced training mechanism trains a generic model that coordinates nodes to optimize network performance in communication scenarios with varying malfunction rates. Algorithm \ref{alg:ATM} outlines the procedures of the advanced training mechanism.
\vspace{-1em}
\begin{algorithm}
\caption{Procedures of the advanced training mechanism}\label{alg:ATM}
\KwIn{Evaluation times $n_{eva}$, learning factor set $\mathcal{T}$, model utility set $\boldsymbol{U}$, utility step size $\triangle u$.}
\KwResult{$(\theta^{*};u^{th},\Gamma)$.}
Prepare the candidate combinations of $u^{th}$ and $\Gamma$ with PE-module.\\
\For{$g \gets 1,\ldots,|\mathcal{T}|$}{
\For{$u^{th} \in [u^{min},u^{max}]$}{
Initialize $u^{th}$ and $\Gamma=\mathcal{T}[g]$ for Algorithm \ref{alg:modelTraining}.\\
\For{$episode \gets 1,2,\ldots,M$}{
\If{current episode is the evaluation episode}{
Evaluate $\bar{\mathcal{U}_{\mathcal{N}}}$ by (\ref{equ:aveUtility}).\\
\If{$\bar{\mathcal{U}_{\mathcal{N}}} \ge u^{th}$}{
Increase $\epsilon$ by (\ref{equ:increEpsilon});\\
}
\Else
{Decrease $\epsilon$ by (\ref{equ:decreEpsilon}).\\
}
}
Execute the trained model and calculate the average reward $\bar{R}(u^{th},\Gamma)$.\\
Store $\bar{R}(u^{th},\Gamma)$ into $\boldsymbol{R}$.\\
}
Update $u^{th} \gets u^{th}+\triangle u$\\
}
Update $g \gets g+1$\\
}
\Return $(\theta^{*};u^{th},\Gamma)=\arg\max \{\boldsymbol{R}|\theta;u^{th},\Gamma\}$
\end{algorithm}
\vspace{-1em}

In the training phase, the utility threshold used to evaluate model performance and the learning factor to control the magnitude of node malfunction rate variation are the most important hyper-parameters of the proposed advanced training mechanism, which are not directly learned within the estimators. Due to the sensitivity of deep reinforcement learning to the setting of algorithmic hyper-parameters, we first construct the candidate combinations of $u^{th}$ and $\Gamma$ for training the ICRL-JSA models.

\subsubsection{Performance Evaluation Module}\label{sec:PE}

Our expectation is that the trained model will have similar utility in IC-UWSNs as in perfect UWSNs. A model for solving FERCOP in UWSNs is initially trained to obtain an upper bound on the network utility $u^{max}$. Then, a completely uncontrolled random power allocation scheme is implemented in order to attain the minimum acceptable network utility value $u^{min}$. The range of utility thresholds is therefore determined, i.e., $u^{th} \!\in\! [u^{min},u^{max}]$. The learning factor set $\mathcal{T}$ is determined empirically, containing $|\mathcal{T}|$ candidates. The feasible domain of $u^{th}$ is discretized into a set containing $s$ elements spaced by $\triangle u$, such that $u^{max}\!=\!u^{min}\!+\!(s\!-\!1) \times \triangle u$. A grid search is performed to find the optimal combination of $u^{th}$ and $\Gamma$ that gives the highest network utility. The PE-module compares the average model performance $\bar{\mathcal{U}_{\mathcal{N}}}$ in IC-UWSNs with the utility threshold only at regular intervals to stabilize the training procedure. $\bar{\mathcal{U}_{\mathcal{N}}}$ is as given in (\ref{equ:aveUtility}).
\begin{equation}
\bar{\mathcal{U}_{\mathcal{N}}}=\frac{1}{n_{eva}}\sum\nolimits_{i=1}^{n_{eva}}\mathcal{U}_{\mathcal{N}}^{i}
\label{equ:aveUtility}
\end{equation}
where $n_{eva}$ is the model evaluation times, $\mathcal{U}_{\mathcal{N}}^{i}$ is the network utility of the $i$-th evaluation. The comparison results are fed into the MRA-module for controlling the direction (positive or negative) of node malfunction rate adjustment. After each training, the PE-module executes the trained model $n_{eva}$ times and calculates the average accumulated reward $\bar{R}(\theta;u^{th},\Gamma)$ as (\ref{equ:aveR}). $\bar{R}(\theta;u^{th},\Gamma)$ is stored in set $\boldsymbol{U}$ that is used to determine the optimal model once all training procedures have been finished.
\begin{equation}
\bar{R}(\theta;u^{th},\Gamma)=\frac{1}{n_{eva}}\sum\nolimits_{i=1}^{n_{eva}}R^{i}
\label{equ:aveR}
\end{equation}
where $R^{i}$ is the accumulated reward of the $i$-th evaluation.

\subsubsection{Node Malfunction Rate Adjustment Module}\label{sec:MRA}

The fundamental idea behind the MRA-module is to regulate training difficulty by adjusting the node malfunction rate at the beginning of each evaluation episode during the training phase. Through the MRA-Module, we aim to progressively prepare nodes to handle scenarios where other collaborators within the system experience malfunctions during the training phase. More specifically, the MRA-module periodically modifies the node's malfunction rate $\epsilon$ in response to the comparison results from the PE-module. $\epsilon$ is assigned a value ranging from 0 to $\epsilon^{max}$. If the model performs better than expected (i.e., $\bar{\mathcal{U}_{\mathcal{N}}} \!\ge\! u^{th}$), the node malfunction rate is increased by (\ref{equ:increEpsilon}) to intensify the training difficulty. Conversely, if the model's performance falls below expectations, $\epsilon$ is decreased by (\ref{equ:decreEpsilon}). With a higher utility threshold, the node is expected to deliver improved performance in the current environment. The central controller sends the updated malfunction rate and model parameters $\theta$ to each ICRL-JSA agent. The nodes then update their malfunction states $\kappa$ in accordance with $\epsilon$ in the subsequent training episode and use the updated $\theta$ to estimate the value function $Q_{\theta}(o_{i},a_{i}), i \!\in\! \mathcal{N}$.
\begin{equation}
\epsilon=\max\{\epsilon+\Gamma(1-\epsilon),\epsilon^{max}\}
\label{equ:increEpsilon}
\end{equation}
\begin{equation}
\epsilon=\min\{0,\epsilon+\Gamma(0-\epsilon)\}
\label{equ:decreEpsilon}
\end{equation}

\vspace{-1em}
\subsection{Practicality of ICRL-JSA}

When considering the implementation of intelligent algorithms in practical underwater systems, concerns may arise regarding 1) the computational energy consumption of the learning-based algorithm, and 2) the difficulties of the fine-tuning phase during deployment that mitigates the gap between simulator and reality. \textit{First}, the deep MARL-based optimization scheme is considered acceptable only if the energy used for decision-making is significantly lower than that required for transmission \cite{li2020routing}. In the proposed deep MARL-based method, each intelligent node, referred to as an ICRL-JSA node, is tasked with performing computations related to the Q-values of available transmit power levels. However, these computations are executed in a direct end-to-end manner, resulting in reduced power consumption that is notably lower than that of acoustic communications (lower by two orders of magnitude) \cite{freitag2005whoi}\cite{hu2010qelar}. Consequently, the computational overhead imposed by the RL-based method is considered acceptable in real-world deployments.

\textit{Second}, due to the challenges in obtaining real-world data, such as sample inefficiency and collection costs, simulation environments are commonly employed for training and evaluating DRL-based decision-making models \cite{casari2020asuna}. However, the effectiveness of models trained in simulators may be compromised by disparities between simulated and real-world conditions. Transferring policies learned from simulation to the real world presents a challenging issue. In addressing the online adaptation issue, some researchers are dedicated to developing more realistic simulators that minimize the mismatch between training data and real-world experiences. This allows RL agents pre-trained in these simulators to be directly deployed in real-life scenarios without requiring additional training or adjustments \cite{zhao2020sim}. While other researchers focus on online transfer strategies that enable pre-trained models to be fine-tuned through real interactions with the system during deployment. Commonly used methods include transfer learning, domain adaptation, and the emerging digital twin method. In this paper, we mitigate the sim-to-real gap by constructing a simulator that reflects the unique characteristics of the acoustic channel to simulate the propagation of acoustic signals. We also consider the energy-constrained and imperfect nature of UWSNs. To enhance the generalization ability of ICRL-JSA, further research is needed to develop effective policy transfer strategies capable of handling variations between simulation and practical applications.

\section{Performance Evaluation}\label{sec:pe}

This section presents the numerical results for ICRL-JSA. We evaluated ICRL-JSA with seven baseline methods and its variants in several representative scenarios. Our experiments address the following questions:
\begin{enumerate}[(1)]
\item How does ICRL-JSA compare to widely used underwater link scheduling and power allocation strategies; and how does ICRL-JSA adapt as communication scenario and scheduling environment changes ($\S$\ref{sec:comBase})?
\item How does each of the main components contribute to the performance of ICRL-JSA ($\S$\ref{sec:admJSA}); and how does the designed \textit{advanced training mechanism} improve network performance in more complicated communication environments ($\S$\ref{sec:admeva})? 
\item How does the model hyper-parameters of ICRL-JSA affect the network performance ($\S$\ref{sec:hype})?
\end{enumerate}

\vspace{-1em}
\subsection{Evaluation Setup}

Consider a UWSN with $N$ underwater nodes and one mobile entity. The transmitters are evenly distributed at the bottom of the cylinder region, and the receivers are located at the top. The cylinder region has a radius of 4 km and a height of 1 km. The mobile node randomly enters and exits the region along a straight path. The mobile node operates at a constant power level of 4 W. At the beginning of each time slot, transmitters can communicate with their designated receivers. Otherwise, they wait for subsequent time slots. Transmission ceases when the remaining energy falls below 10 percent. The adaptive fairness evaluation horizon $h\!=\!|\mathcal{N}_{send}|$, the normalization constant $A_{0}$ is 1, and the geometric spreading factor $k$ is 1.5. Mobility model parameters follow those in \cite{he2020trust}. Regarding application requirements, $T_{tran}$ is set to 3 seconds in both scenarios, $T_{guard}$ is 0.1 second by referring to \cite{song2019optimizing}, and $T_{prop}$ is determined using the BELLHOP method. In the evaluations, we assign equal significance to spatial reuse, communication fairness, and delivery ratio. Therefore, the exemplary weight parameters $\alpha$, $\beta$, and $\mu$ for the long-term network utility, specifically for spatial reuse utility, communication fairness utility, and ineffective communication utility are set to be $\{1,1,1\}$. The SINR threshold can also be adjusted based on specific application requirements and communication environment without affecting the effectiveness of the proposed method. The training and execution of ICRL-JSA focus on successful reception rather than exact signal strength. Table \ref{tab:paraset} lists the simulation parameters.
\vspace{-1em}
\begin{table}[htp]
\caption{Parameters.}
\vspace{-1em}
\begin{center}
\begin{tabular}{ll}
\hline
Parameter&Value\\
\hline
Transmit power $p$&[2, 4, 8, 16, 32, 64] W\\
Battery capacity $E^{0}$&5,000 J\\
Transducer efficiency $\eta_{0}$&100\%\\
Network lifetime requirement $\delta^{0}$&30 time slots\\
Carrier frequency $f$&8 kHz\\
Bandwidth $B$&3 kHz\\
Communication threshold $\gamma^{th}$&10 dB\\
\hline
\end{tabular}
\end{center}
\label{tab:paraset}
\end{table}
\vspace{-1em}

We implemented a five-layer deep neural network for the ICRL-JSA agents, comprising an input layer, two fully connected (FC) layers, one GRU, and an output layer. Each FC layer consists of 64 neurons, and the two FC layers are connected by a GRU with 64 hidden units. The output layer provides the Q-values for each possible action. The activation function for all neurons is ReLU. The discount factor $\gamma$ is set to 0.99. The size of the experience buffer $B_{r}$ and mini-batch size $b$ are configured as 10,000 and 32, respectively. All networks use the ADAM optimizer, and the learning rate is 0.0005. The model was trained with 200,000 episodes. At every 200 episodes, we assess the current model through 20 runs, averaging the mean episode rewards. Evaluated models are stored, contributing to the determination of the optimal model. To balance exploration and exploitation during training, we employ an $\varepsilon$-greedy behavior policy, with the greedy factor $\varepsilon$ declining linearly from 1 to 0.05 over the initial 100,000 episodes and remaining constant at $\varepsilon=0.05$ for the remaining training episodes. 

In this paper, all training and simulations are conducted in Python, utilizing the PyTorch library for DRL implementation. The propagation of acoustic signal is simulated through arlpy simulator \cite{arlpy2023}, which serves as a Python interface to BELLHOP. The simulations are carried out on a computer equipped with a GeForce RTX 3070Ti GPU, an Intel i9-12900H CPU @ 2.5GHz, and 32GB RAM.

\vspace{-1em}
\subsection{Comparison with Baseline Methods}\label{sec:comBase}

\begin{figure*}[htbp]
\begin{center}
\includegraphics[width=6.8in]{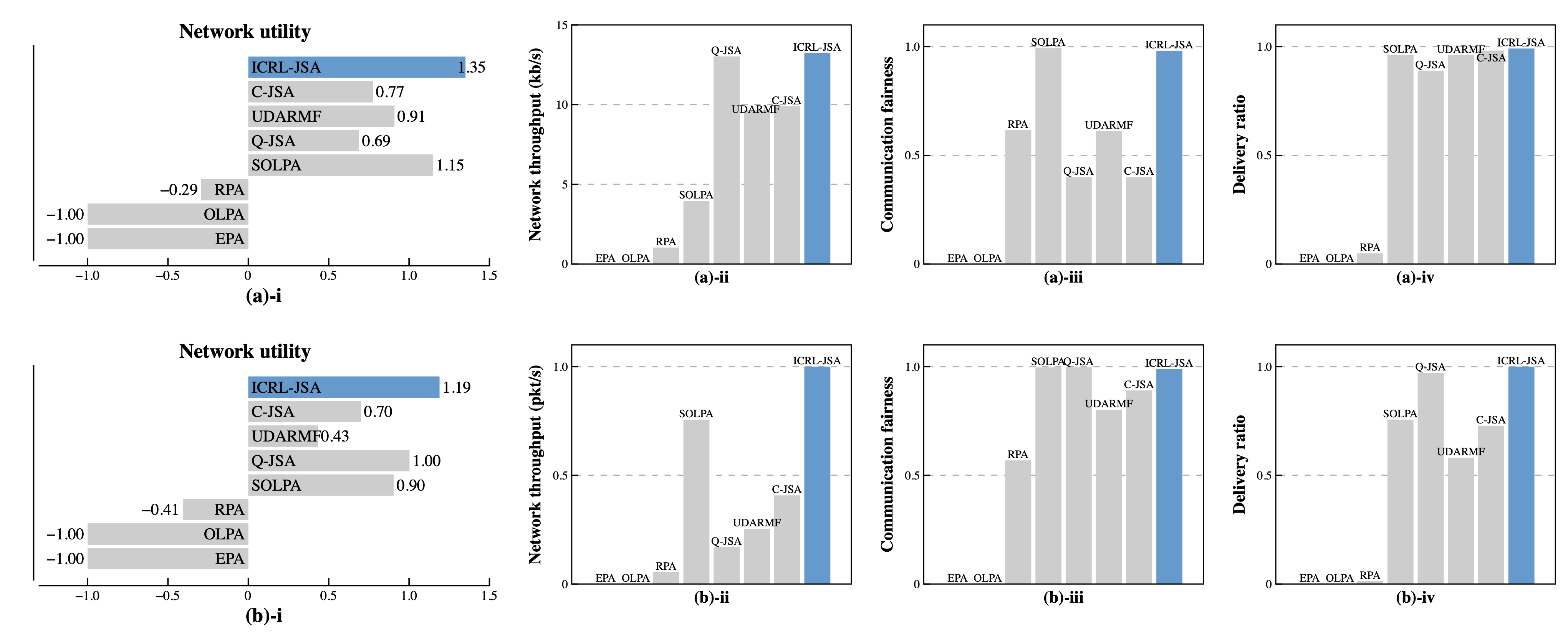}
\vspace{-1em}
\caption{Effects of transmission strategy on network utility in UWSNs. (a) Unicast scenario, (b) Broadcast scenario.}
\label{fig:noUtility}
\end{center}
\end{figure*}

In this section, we evaluate the effectiveness of ICRL-JSA in two representative communication scenarios, which are the UWSNs and IC-UWSNs communication scenario, and compare it to the following baselines: (1) \textit{Equal power allocation (EPA)}\cite{yu2018power}: All nodes transmit simultaneously and operate identically at the maximum available power; (2) \textit{Open-loop power allocation (OLPA)}\cite{pompili2009cdma}: All nodes transmit simultaneously while adapting their power levels based on the propagation model; (3) \textit{Random power allocation (RPA)}\cite{lee2011distributed}: Nodes transmit at arbitrary power levels and are allowed to refrain from transmitting during some time slots; (4) \textit{Slotted open-loop power allocation (SOLPA)}: Nodes select their transmit power at each allocated slot according to the propagation model; (5) \textit{Q-learning-based joint link scheduling and power allocation (Q-JSA)}\cite{wang2019self}: A Q-learning-based joint link scheduling and power allocation to alleviate interference in underwater communications, thereby improving network capacity; (6) \textit{UDARMF}\cite{zhang2021udarmf}: At each allocated slot, the nodes cooperatively select their transmit power to maximize network capacity under energy constraints; (7) \textit{COMA-based joint link scheduling and power allocation (C-JSA)}\cite{foerster2018counterfactual}: At each time slot, the nodes adjust their transmission parameters based on the counterfactual multi-agent (COMA) policy gradients. Table \ref{tab:JSAbase} summarizes the baseline methods.
\begin{table}[htp]
\caption{Baseline methods for comparing the effectiveness of ICRL-JSA}
\vspace{-1em}
\begin{center}
\begin{tabular}{lll}
\hline
Method&Link scheduling&Power allocation\\
\hline
EPA&/&Maximum available\\
RPA&Random&Random\\
OLPA&/&Propagation model-based\\
SOLPA&Allocated slot&Propagation model-based\\
Q-JSA&Learning-based&Learning-based\\
UDARMF&Learning-based&Learning-based\\
C-JSA&Learning-based&Learning-based\\
\hline
\end{tabular}
\end{center}
\label{tab:JSAbase}
\end{table}
\vspace{-1em}

The propagation model-based power allocation is described in \cite[Ch. 15]{dhanak2016springer} and \cite{pompili2009cdma}, which maps the mathematical transmit power into the acoustic modem design. It should be noted that the executed transmit power $p_{e}$ is constrained within the range $[p_{min},p_{max}]$. And according to the acoustic modem design as well as the battery capacity, the maximum available transmit power for all EPA nodes in the considered system is 32 W.

\subsubsection{Evaluations in UWSNs}

Figure \ref{fig:noUtility} illustrates the impact of various transmission strategies on network utility, throughput, communication fairness, and delivery ratio in UWSNs without node malfunctions. The considered scenarios involve unicast communication and broadcast scenario, sharing identical network deployments. The unicast scenario consists of five transmitter-receiver pairs, while the broadcast scenario comprises ten underwater nodes, with five of them serving as broadcast transmitters. As depicted in Fig. \ref{fig:noUtility}(a)-i, ICRL-JSA attains the highest network utility, with learning-based methods outperforming non-learning-based baselines on average. Among the non-learning-based baselines, SOLPA exhibits higher network utility than RPA, OLPA, and EPA. The superiority of SOLPA mainly stems from its alternative transmission mechanism, providing it with higher communication fairness and delivery ratio. However, due to its limited utilization of channel resources, its throughput is significantly lower compared to ICRL-JSA and other learning-based methods.

In the broadcast scenario, the network faces increased communication interference, leading to greater optimization challenges. ICRL-JSA still achieves the highest network utility. Among all learning-based approaches, UDARMF, designed to optimize network capacity, exhibits the most significant performance degradation, particularly in throughput and delivery ratio. Regarding Q-JSA, it demonstrates enhanced communication fairness and delivery ratio in the broadcast scenario, but the network throughput is lags behind other intelligent methods. Through a further investigation of its transmission behavior, we observed that Q-JSA nodes transmit only in a few time slots, with little to no concurrent communication. These observations emphasize the necessity of tailoring reward functions to specific application requirements. Additionally, ICRL-JSA consistently maintains a delivery ratio close to 1 in both scenarios, indicating its suitability for energy-constrained UWSNs.

We also incorporate evaluations and discussions comparing the proposed ICRL-JSA with the centralized solution in unicast scenario. Notably, ICRL-JSA formulates decisions using local observations, bridging the partial-global gap by leveraging GRU. In contrast, the centralized approach assumes an additional central node that gathers the observation information from all communication links within the network, and distributes the transmission parameters to the underwater nodes. After executing the model for 100 times and calculating the average, ICRL-JSA and the centralized approach exhibit comparable performance. The network utility is 1.351 and 1.352, the throughput is 13.237 kb/s and 6.371 kb/s, and the delivery ratio is 0.991 and 0.998, respectively. Both methods achieve a communication fairness of 0.982.

While the centralized approach demonstrates superior delivery ratio compared to ICRL-JSA, real-world deployments may introduce performance degradation, particularly in network throughput, for several reasons. Firstly, the process of collecting observations at each transmitter consumes a substantial amount of energy and bandwidth, ultimately resulting in a shortened network lifetime and reduced communication efficiency. Secondly, the centralized approach requires full observations of all transmitters, implying that the central node (if existed) requires more memory to store these information. As the network scales up, the storage burden proportionally increases. Moreover, when the decision-making model relies on an extensive amount of input information, the training process becomes more intricate. Consequently, future research should prioritize finding the balance between addressing the partial-global gap in distributed approaches and managing the communication, storage, and computation overhead associated with centralized methods.

\begin{figure*}[htbp]
\begin{center}
\includegraphics[width=6.8in]{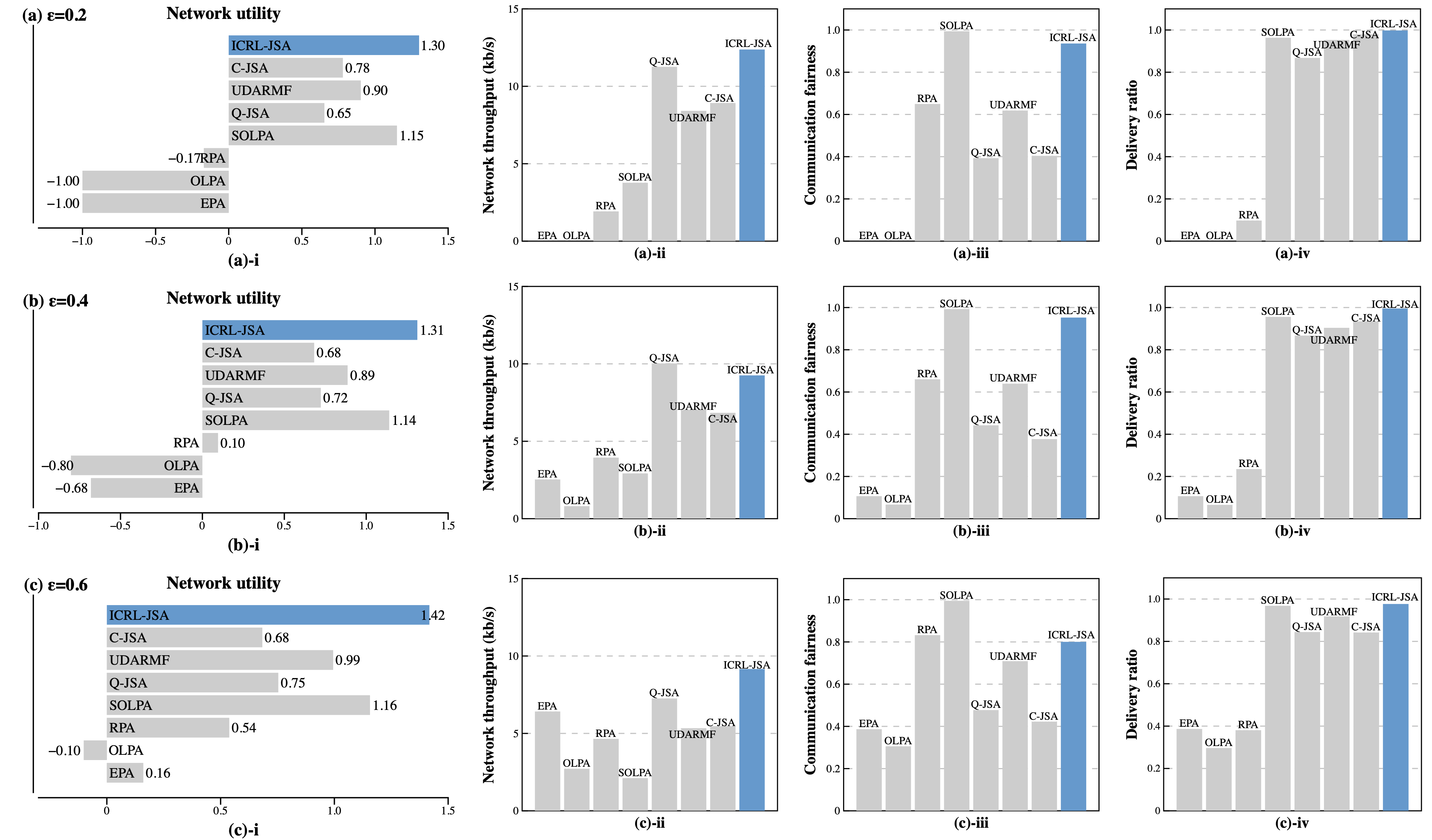}
\vspace{-1em}
\caption{Effects of transmission strategy on network performance in IC-UWSNs. (a) $\epsilon=0.2$, (b) $\epsilon=0.4$, (c) $\epsilon=0.6$.}
\label{fig:crashutility}
\end{center}
\end{figure*}

\subsubsection{Evaluations in IC-UWSNs}

We also make evaluations to compare the performance of ICRL-JSA and baseline methods in IC-UWSNs with five transmitter-receiver pairs. Figure \ref{fig:crashutility} illustrates how different node malfunction rates result in vastly different network performance. There are three key observations from the results. \textit{First}, although EPA and OLPA fail the network when $\epsilon$=0 and $\epsilon$=0.2, as shown in Fig. \ref{fig:noUtility}(a) and Fig. \ref{fig:crashutility}, they achieve superior performance with greater malfunction rates. This is because both EPA and OLPA lack link scheduling strategies and transmit at a constant power level. As more nodes degrade into random nodes, the number of concurrent communications decreases, allowing for increased successful communications and enhanced network performance. Furthermore, SOLPA gradually loses its effectiveness as $\epsilon$ increases. This is attributed to SOLPA's alternative transmission mechanism, which fails to fully utilize the communication resources released by malfunction nodes, resulting in a decrease in network throughput. These results demonstrate the effectiveness of link scheduling and power allocation strategies in dense networks.

\textit{Second}, improving network throughput is at the expense of communication fairness and vice versa. In various scenarios, ICRL-JSA investigates the optimal trade-off design between spatial reuse and communication fairness while increasing network reliability by raising the delivery ratio, i.e., to maximize network utility. In contrast, other learning-based methods, such as Q-JSA that aims to optimize network throughput, sacrificing fairness for higher network throughput, resulting in lower network utility.

\textit{Third}, ICRL-JSA shines in more complex scheduling environments with higher malfunction rates, where scheduling decisions have a much more significant impact when the nodes are fully-cooperative. Recall that ICRL-JSA is trained on IC-UWSNs, while the other learning-based baseline methods are trained on perfect UWSNs. When more intelligent nodes degrade to malfunction nodes, UDARMF, C-JSA, and Q-JSA experience significant degradations in network throughput, indicating that it is necessary to consider the practical issues of real-world systems when designing intelligent optimization algorithms.

\subsection{Impact of ICRL-JSA Architecture}\label{sec:admJSA}

\begin{table*}[htp]
\caption{Methods for comparing the effectiveness of the ICRL-JSA architecture.}
\vspace{-1em}
\begin{center}
\begin{tabular}{lllll}
\hline
Method&Fairness&Link scheduling&Power allocation&Delivery\\
\hline
ICRL-PA (LH)&Long-term&/&Learning-based&$\surd$\\
ICRL-PA (AH)&Adaptive&/&Learning-based&$\surd$\\
ICRL-LS (LH)&Long-term&Learning-based&Propagation model-based&$\surd$\\
ICRL-LS (AH)&Adaptive&Learning-based&Propagation model-based&$\surd$\\
ICRL-JSA (LH, w/o eff)&Long-term&Learning-based&Learning-based&/\\
ICRL-JSA (LH)&Long-term&Learning-based&Learning-based&$\surd$\\
\hline
\end{tabular}
\end{center}
\label{tab:JSAarch}
\end{table*}
\begin{figure*}[htbp]
\begin{center}
\includegraphics[width=5.9in]{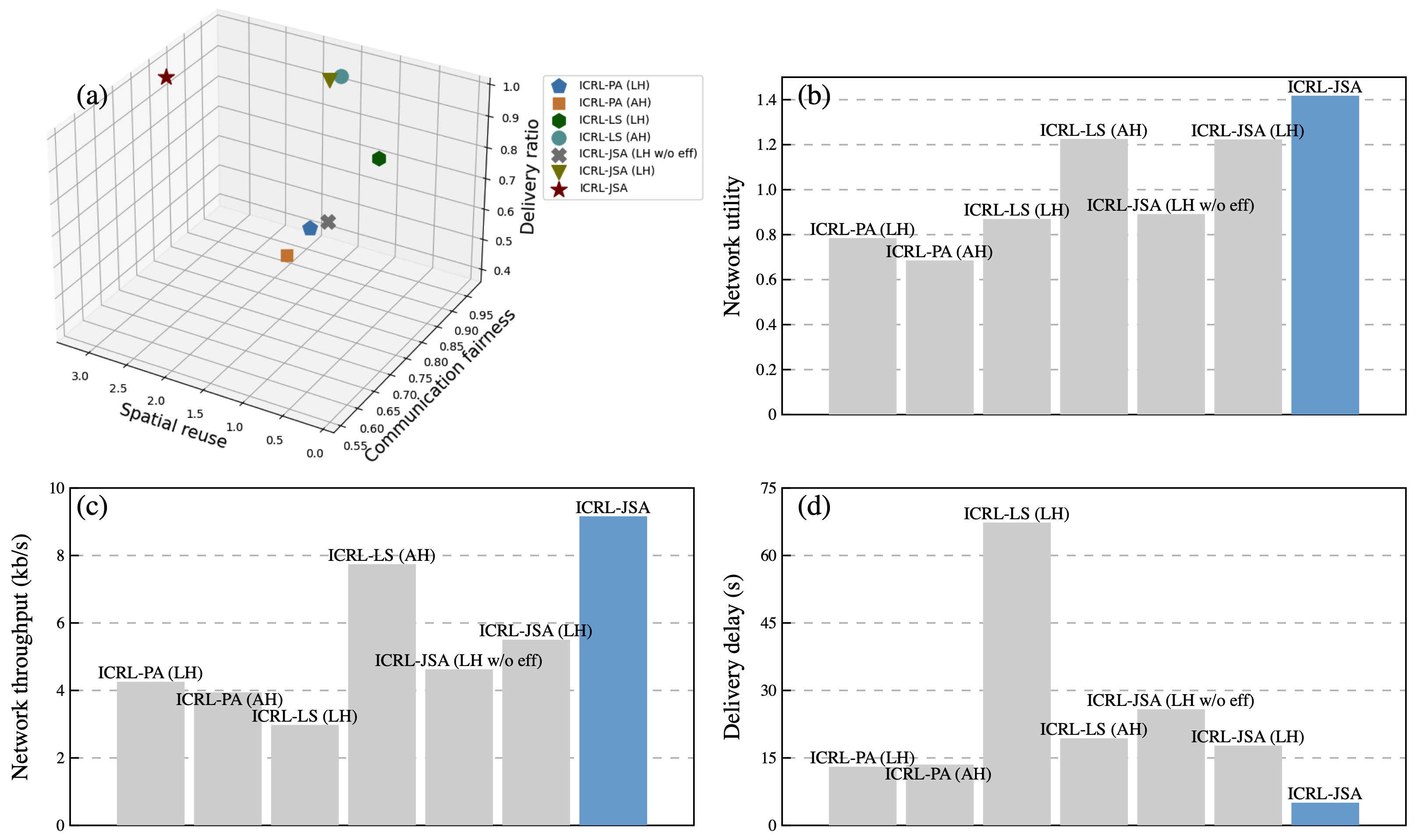}
\vspace{-1em}
\caption{Effects of the ICRL-JSA architecture on network performance in IC-UWSNs ($|\mathcal{N}_{send}|$=5, $\epsilon=0.6$). (a) Optimization objectives, (b) network utility, (c) network throughput, and (d) delivery ratio.}
\label{fig:crash02}
\end{center}
\end{figure*}

We validate that ICRL-JSA requires all of its key design components by selectively omitting certain components. We train six variants of ICRL-JSA, and compare them to ICRL-JSA in IC-UWSNs ($|\mathcal{N}_{send}|$=5, $\epsilon$=0.6). The variants are compared in Table \ref{tab:JSAarch}. Specifically, \textit{LH} refers to the scenario where the fairness evaluation horizon $h$ in (\ref{equ:EFRAH}) is $t$, \textit{w/o eff} refers to the circumstances where the penalty term $\varsigma$ in (\ref{equ:EFRAH}) is omitted.

Figure \ref{fig:crash02} demonstrates that in IC-UWSNs, omitting each key idea from ICRL-JSA results in lower network utility and worse network performance than the fine-tuned resource allocation model. The three dimensions presented in Fig. \ref{fig:crash02}(a) refer to the three optimization objectives of the formulated problem, i.e., spatial reuse ($x$-axis), communication fairness ($y$-axis), and delivery ratio ($z$-axis). Each data point represents the optimization objective values of the solution obtained by ICRL-JSA and its six variants, and we can find the superiority relationship of these methods from the figure. Since we aim to minimize the number of ineffective communications and its calculation yields a negative result as (\ref{equ:Ief}). We use the positive delivery ratio to illustrate the negative ineffective communications for the sake of simplicity. Spatial reuse refers to the average number of concurrent communications in the network throughout the network lifetime. It is evident from Fig. \ref{fig:crash02}(b) that ICRL-JSA has greater network utility than other variants.

We then dive deeper to further illustrate the impact of our key designs on network performance, yielding three notable observations. \textbf{First}, the \textit{joint link scheduling and power allocation scheme} has the most significant impact on ICRL-JSA's performance. Absence of link scheduling results in simultaneous transmission by all nodes, determining transmit power based on partial local observations through reinforcement learning. Despite ICRL-PA (LH) and ICRL-PA (AH) achieving enhanced spatial reuse, their network utility is lower than other models (ranging from 9.76\% to 51.64\%, making FER communications challenging in IC-UWSNs. Without power allocation, communication fairness is hard to maintain. The two ICRL-LS methods (especially the LH one) accumulate significant delivery delays, and the spatial reuse is the lowest among all methods. \textbf{Second}, neglecting the \textit{delivery guarantee mechanism} (i.e., omit the penalty term $\varsigma$) drastically reduces the delivery ratio of ICRL-JSA, diminishing network throughput and increasing delivery delays. \textbf{Third}, models employing an \textit{adaptive fairness evaluation horizon} during the training phase outperform those with a lifetime fairness evaluation horizon on average. This suggests that a shorter evaluation horizon allows for more nodes to access the acoustic channel during short intervals, thereby enhancing network performance.

\subsection{Evaluation of the Advanced Training Mechanism}\label{sec:admeva}

In this section, we seek to validate the superiority of our designed \textit{advanced training mechanism} across three scheduling environments (i.e., $\epsilon$=0.2, $\epsilon$=0.4, $\epsilon$=0.6) through comparison with three alternative training mechanisms as follows: (1) \textit{RL-JSA:} Models are trained without considering node malfunctions (i.e., pm=0). (2) \textit{ICRL-JSA with predefined malfunction rate (ICRL-JSA w/pm):} Models are trained in imperfect environments with predefined malfunction rates (i.e., pm=0.2, pm=0.4, pm=0.6). (3) \textit{ICRL-JSA with incremental malfunction rate (ICRL-JSA w/im):} Models are trained in imperfect environments with an incremental malfunction rate (i.e., $im\in[0,0.6]$).
\begin{figure}[htbp]
\begin{center}
\includegraphics[width=3in]{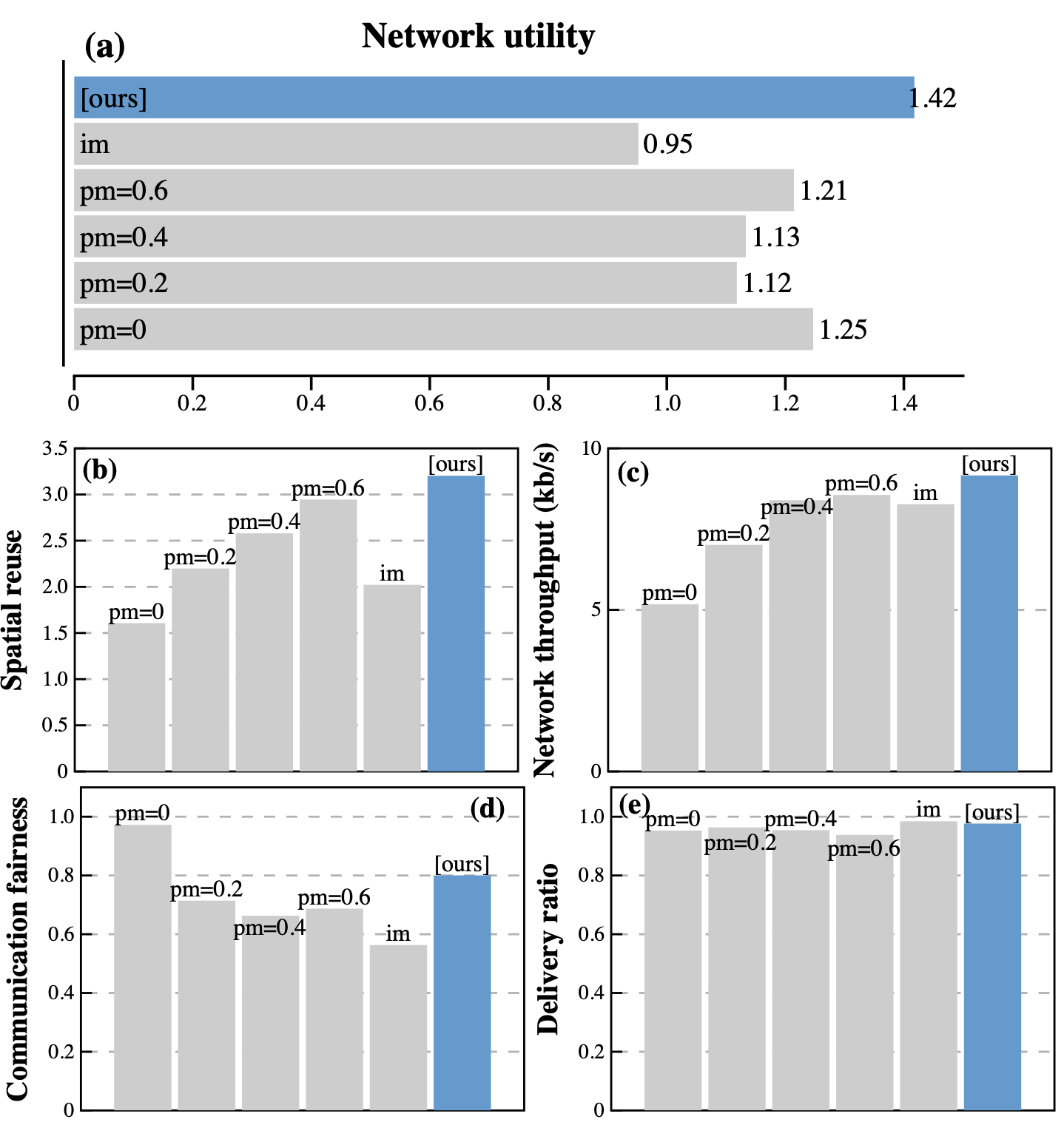}
\vspace{-1em}
\caption{Performance of the advanced training mechanism and other baseline training mechanisms in IC-UWSNs with $\epsilon=0.6$. (a) Network utility, (b) spatial reuse, (c) network throughput, (d) communication fairness, (e) delivery ratio.}
\label{fig:robustLearning}
\end{center}
\end{figure}

Figure \ref{fig:robustLearning} demonstrates the effectiveness of the designed advanced training mechanism in addressing communication scenarios with node malfunctions. In the current scenario, ICRL-JSA attains a spatial reuse at 3.20, surpassing other training mechanisms by 8.75\% to 99.31\% (from pm=0.6 to pm=0). Despite being trained in a perfect environment, pm=0 exhibits superior network utility compared to other baseline methods. This result emphasizes the importance of adequately capturing the characteristics of UWSNs during model training. Other training environments with node malfunctions may not adequately reflect the features of UWSNs. Additionally, although the network utility is comparable between pm=0 and pm=0.6, pm=0.6 optimizes spatial reuse and throughput at the expense of communication fairness. In contrast, pm=0 better maintains communication fairness but performs poorly in the other two aspects. In comparison, ICRL-JSA achieves a balance among multiple network performance metrics. Furthermore, the performance of im is consistently inferior in all evaluations, possibly due to the frequent changes in the training environment, preventing the model from thoroughly exploring the communication scenario. The proposed MRA-module in this paper demonstrates greater rationale.

\subsection{Impact of ICRL-JSA Hyper-Parameters on Network Performance}\label{sec:hype}

\begin{figure}[htbp]
\begin{center}
\includegraphics[width=3.1in]{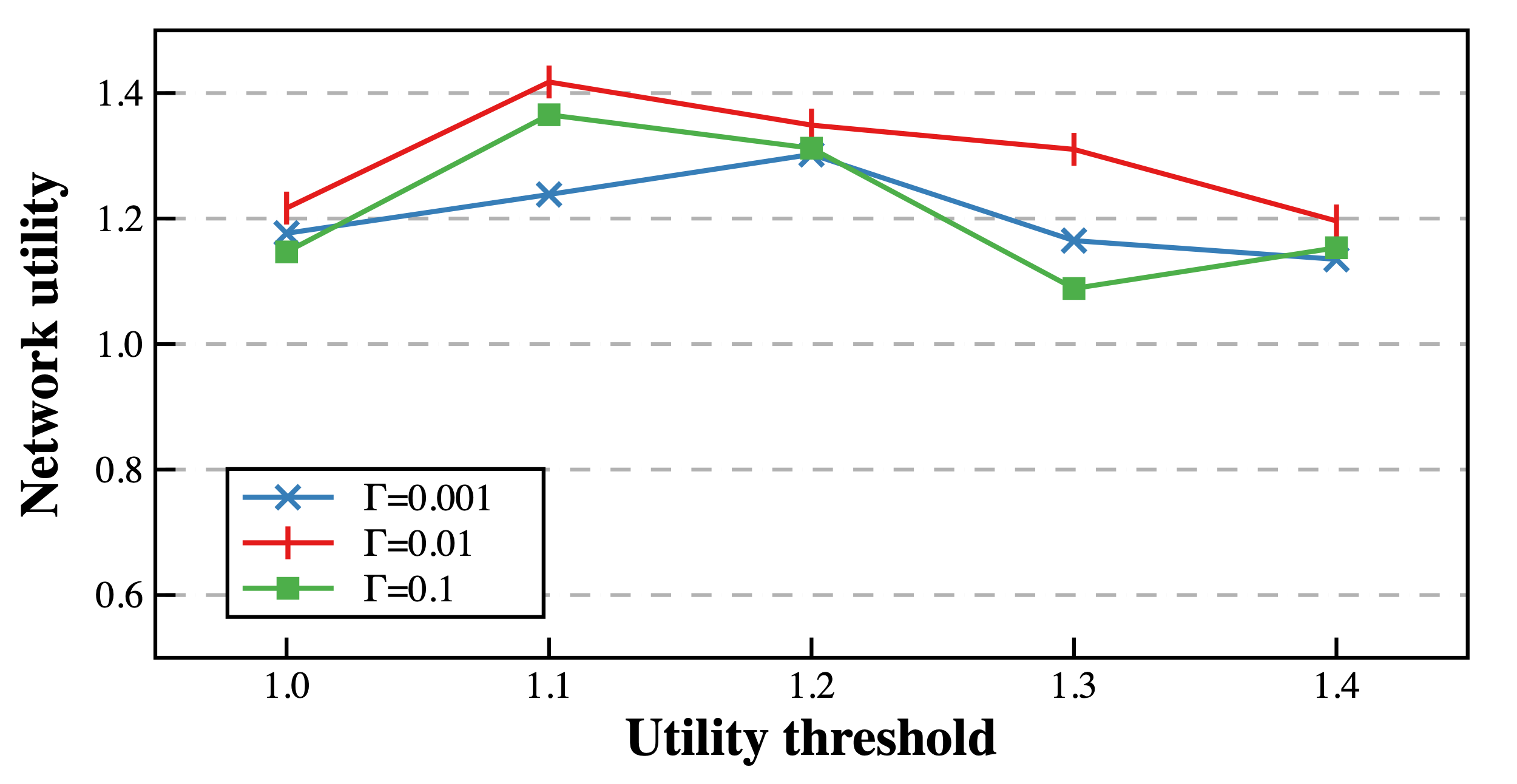}
\vspace{-1em}
\caption{Effects of utility threshold and learning factor on network utility.}
\label{fig:hpUtility02}
\end{center}
\end{figure}

The effect of network utility threshold $u^{th}$ and learning factor $\Gamma$ on network utility is depicted in Fig. \ref{fig:hpUtility02}. As explained in $\S$\ref{sec:ATM}, $u^{th}$ evaluates the model performance, and $\Gamma$ controls the magnitude of node malfunction rate variation. During the training process, ICRL-JSA increases the environment malfunction rate if $\mathcal{U}_{\mathcal{N}}$ meets $u^{th}$; otherwise, ICRL-JSA decreases the environment malfunction rate until the model reaches the utility threshold or the training is complete.

The model exhibits enhanced network utility when $u^{th}$  is set to 1.1 and $\Gamma$ is set to 0.01. The performance of the model varies with the three learning factors, i.e., $\Gamma$=0.001, $\Gamma$=0.01, and $\Gamma$=0.1. For $\Gamma$=0.1, the network utility obtained by the model fluctuates between 1.0 and 1.4 as the utility threshold increases. In cases where $\Gamma$=0.001 and $\Gamma$=0.01, the network utility experiences an initial increase and subsequent decrease as the utility threshold increases. Models with $\Gamma$=0.01 consistently outperform those with $\Gamma$=0.001.

Two conclusions can be drawn. First, determining the optimal model necessitates an appropriately chosen learning factor $\Gamma$. The performance inflection point of the model cannot be effectively identified with either a rapid or sluggish learning pace. Second, a gracefully adjustment of the training threshold $u^{th}$ can motivate the model to achieve higher performance. Excessive thresholds may lead to models consistently trained with a zero malfunction rate, rendering them unable to cope with progressively complex scenarios. This paper empirically derives the candidate values for the learning factor, and future research will focus on the adaptive selection of the hyper-parameters.

\section{Conclusions and Future Work}\label{sec:con}

In this paper, we address the fair, efficient, and reliable communication optimization problem (FERCOP) within the context of IC-UWSNs. We introduced ICRL-JSA, a deep MARL-based joint link scheduling and power allocation scheme for IC-UWSNs. ICRL-JSA provides solutions to FERCOP, contributing to improved communication fairness and delivery ratios. The algorithm dynamically determines how many transmitters can transmit in the current communication environment and how the scheduled nodes will transmit to meet the communication threshold without suppressing other concurrent transmissions. Simulation results have validated the effectiveness of ICRL-JSA in providing fair, efficient, and reliable communication across various scenarios. The design components and advanced training mechanism employed in ICRL-JSA have potential applicability to diverse underwater tasks, provided that energy supply, acoustic channel characteristics, and unexpected node malfunctions still limit the performance of UWSNs.

The ICRL-JSA algorithm, while demonstrating notable performance in IC-UWSNs, exhibits certain limitations that will be addressed in our future research. First, the learning factor of the advanced training mechanism is currently predetermined. A learned or adaptive step size will be the subject of future research. Second, despite its superior adaptability and robustness compared to existing schemes, enhancing the generalization capabilities of ICRL-JSA will be a specific focus for improvement. Third, we will involve the integration of modulation techniques and more precise SINR threshold estimation methods into the strategy to provide more practical solutions for UWSNs.

 \section*{Acknowledgments}

This work was supported by the National Key Research and Development Program (Grant No. 2021YFC2803000), the National Key Basic Research Program (Grant No. 2020YFB050001), the National Natural Science Foundation of China (Grants No. 62101211 and No. 61971206), and the Joint Funds of the National Natural Science Foundation of China (Grant No. U22A2009).

\ifCLASSOPTIONcaptionsoff
  \newpage
\fi



%
\bibliographystyle{IEEEtran}
\bibliography{IEEEabrv,mybibfile}

\begin{thebibliography}{10}
\providecommand{\url}[1]{#1}
\csname url@samestyle\endcsname
\providecommand{\newblock}{\relax}
\providecommand{\bibinfo}[2]{#2}
\providecommand{\BIBentrySTDinterwordspacing}{\spaceskip=0pt\relax}
\providecommand{\BIBentryALTinterwordstretchfactor}{4}
\providecommand{\BIBentryALTinterwordspacing}{\spaceskip=\fontdimen2\font plus
\BIBentryALTinterwordstretchfactor\fontdimen3\font minus
  \fontdimen4\font\relax}
\providecommand{\BIBforeignlanguage}[2]{{%
\expandafter\ifx\csname l@#1\endcsname\relax
\typeout{** WARNING: IEEEtran.bst: No hyphenation pattern has been}%
\typeout{** loaded for the language `#1'. Using the pattern for}%
\typeout{** the default language instead.}%
\else
\language=\csname l@#1\endcsname
\fi
#2}}
\providecommand{\BIBdecl}{\relax}
\BIBdecl

\bibitem{satake2014advances}
K.~Satake, ``Advances in earthquake and tsunami sciences and disaster risk
  reduction since the 2004 indian ocean tsunami,'' \emph{Geoscience Letters},
  vol.~1, no.~1, pp. 1--13, 2014.

\bibitem{aalsalem2018wireless}
M.~Y. Aalsalem, W.~Z. Khan, W.~Gharibi, M.~K. Khan, and Q.~Arshad, ``Wireless
  sensor networks in oil and gas industry: Recent advances, taxonomy,
  requirements, and open challenges,'' \emph{Journal of network and computer
  applications}, vol. 113, pp. 87--97, 2018.

\bibitem{xu2016digital}
L.~Xu and T.~Xu, \emph{Digital Underwater Acoustic Communications}.\hskip 1em
  plus 0.5em minus 0.4em\relax Academic Press, 2016.

\bibitem{stojanovic2009underwater}
M.~Stojanovic and J.~Preisig, ``Underwater acoustic communication channels:
  Propagation models and statistical characterization,'' \emph{IEEE
  communications magazine}, vol.~47, no.~1, pp. 84--89, 2009.

\bibitem{gou2022DMPM}
Y.~Gou, T.~Zhang, T.~Yang, J.~Liu, S.~Song, and J.-H. Cui, ``A deep marl-based
  power-management strategy for improving the fair reuse of uwsns,'' \emph{IEEE
  Internet of Things Journal}, vol.~10, no.~7, pp. 6507--6522, 2023.

\bibitem{gupta2000capacity}
P.~Gupta and P.~R. Kumar, ``The capacity of wireless networks,'' \emph{IEEE
  Transactions on information theory}, vol.~46, no.~2, p.~22, 2000.

\bibitem{sathiaseelan2007multimedia}
A.~Sathiaseelan and G.~Fairhurst, ``Multimedia congestion control for broadband
  wireless networks,'' in \emph{2007 16th IST Mobile and Wireless
  Communications Summit}.\hskip 1em plus 0.5em minus 0.4em\relax IEEE, 2007,
  pp. 1--5.

\bibitem{diamant2016leveraging}
R.~Diamant, P.~Casari, F.~Campagnaro, and M.~Zorzi, ``Leveraging the near--far
  effect for improved spatial-reuse scheduling in underwater acoustic
  networks,'' \emph{IEEE Transactions on Wireless Communications}, vol.~16,
  no.~3, pp. 1480--1493, 2016.

\bibitem{gou2021achieving}
Y.~Gou, T.~Zhang, J.~Liu, T.~Yang, S.~Song, and J.-H. Cui, ``Achieving
  time-sharing and spatial-reuse underwater wireless sensor networks with
  communication fairness: A distributed deep multi-agent reinforcement learning
  approach,'' in \emph{The 15th International Conference on Underwater Networks
  \& Systems}, 2021, pp. 1--5.

\bibitem{zhang2021udarmf}
T.~Zhang, Y.~Gou, J.~Liu, T.~Yang, and J.-H. Cui, ``Udarmf: An underwater
  distributed and adaptive resource management framework,'' \emph{IEEE Internet
  of Things Journal}, vol.~9, no.~10, pp. 7196--7210, 2022.

\bibitem{elbatt2004joint}
T.~ElBatt and A.~Ephremides, ``Joint scheduling and power control for wireless
  ad hoc networks,'' \emph{IEEE Transactions on Wireless communications},
  vol.~3, no.~1, pp. 74--85, 2004.

\bibitem{cao2018machine}
X.~Cao, R.~Ma, L.~Liu, H.~Shi, Y.~Cheng, and C.~Sun, ``A machine learning-based
  algorithm for joint scheduling and power control in wireless networks,''
  \emph{IEEE Internet of Things Journal}, vol.~5, no.~6, pp. 4308--4318, 2018.

\bibitem{park2016learning}
T.~Park, N.~Abuzainab, and W.~Saad, ``Learning how to communicate in the
  internet of things: Finite resources and heterogeneity,'' \emph{IEEE Access},
  vol.~4, pp. 7063--7073, 2016.

\bibitem{naderializadeh2021resource}
N.~Naderializadeh, J.~J. Sydir, M.~Simsek, and H.~Nikopour, ``Resource
  management in wireless networks via multi-agent deep reinforcement
  learning,'' \emph{IEEE Transactions on Wireless Communications}, vol.~20,
  no.~6, pp. 3507--3523, 2021.

\bibitem{gorma2019adaptive}
W.~Gorma and P.~D. Mitchell, ``An adaptive tdma-based mac protocol for
  underwater acoustic sensor networks,'' in \emph{Proceedings of the
  International Conference on Underwater Networks \& Systems}, 2019, pp. 1--8.

\bibitem{wang2021concurrent}
Y.~Wang, C.~Zhan, X.~Song, and L.~Lei, ``A concurrent mac protocol with
  master-slave transmission for multi-hop underwater acoustic sensor
  networks,'' in \emph{2021 IEEE 6th International Conference on Signal and
  Image Processing (ICSIP)}.\hskip 1em plus 0.5em minus 0.4em\relax IEEE, 2021,
  pp. 1204--1209.

\bibitem{fan2020link}
Z.~Fan, L.~Wang, B.~Lu, Y.~Yu, C.~Lin, Z.~Luo, Z.~Qin, and M.~Zhu, ``A link
  scheduling algorithm for underwater optical wireless networks,'' in
  \emph{2020 IFIP Networking Conference (Networking)}.\hskip 1em plus 0.5em
  minus 0.4em\relax IEEE, 2020, pp. 827--832.

\bibitem{zhang2021load}
W.~Zhang, X.~Wang, G.~Han, Y.~Peng, M.~Guizani, and J.~Sun, ``A load-adaptive
  fair access protocol for mac in underwater acoustic sensor networks,''
  \emph{Journal of Network and Computer Applications}, vol. 173, p. 102867,
  2021.

\bibitem{islam2022survey}
K.~Y. Islam, I.~Ahmad, D.~Habibi, and A.~Waqar, ``A survey on energy efficiency
  in underwater wireless communications,'' \emph{Journal of Network and
  Computer Applications}, vol. 198, p. 103295, 2022.

\bibitem{sun2023bargain}
Z.~Sun, G.~Sun, Y.~Liu, J.~Wang, and D.~Cao, ``Bargain-match: A game
  theoretical approach for resource allocation and task offloading in vehicular
  edge computing networks,'' \emph{IEEE Transactions on Mobile Computing},
  vol.~23, no.~2, pp. 1655--1673, 2024.

\bibitem{jornet2010joint}
J.~M. Jornet, M.~Stojanovic, and M.~Zorzi, ``On joint frequency and power
  allocation in a cross-layer protocol for underwater acoustic networks,''
  \emph{IEEE Journal of Oceanic engineering}, vol.~35, no.~4, pp. 936--947,
  2010.

\bibitem{yu2018power}
W.~Yu, Y.~Chen, Y.~Tang, and X.~Xu, ``Power allocation for underwater source
  nodes in uwa cooperative networks,'' in \emph{2018 IEEE international
  conference on signal processing, communications and computing
  (ICSPCC)}.\hskip 1em plus 0.5em minus 0.4em\relax IEEE, 2018, pp. 1--6.

\bibitem{zhang2021scalable}
T.~Zhang, Y.~Gou, J.~Liu, T.~Yang, S.~Song, and J.-H. Cui, ``A scalable and
  fair power allocation scheme based on deep multi-agent reinforcement learning
  in underwater wireless sensor networks,'' in \emph{The 15th International
  Conference on Underwater Networks \& Systems}, 2021, pp. 1--5.

\bibitem{cruz2003optimal}
R.~L. Cruz and A.~V. Santhanam, ``Optimal routing, link scheduling and power
  control in multihop wireless networks,'' in \emph{IEEE INFOCOM 2003.
  Twenty-second Annual Joint Conference of the IEEE Computer and Communications
  Societies (IEEE Cat. No. 03CH37428)}, vol.~1.\hskip 1em plus 0.5em minus
  0.4em\relax IEEE, 2003, pp. 702--711.

\bibitem{le2014joint}
A.-M. Le and D.-S. Kim, ``Joint channel and power allocation for underwater
  cognitive acoustic networks,'' in \emph{2014 International Conference on
  Advanced Technologies for Communications (ATC 2014)}.\hskip 1em plus 0.5em
  minus 0.4em\relax IEEE, 2014, pp. 745--749.

\bibitem{wang2022joint}
C.~Wang, W.~Zhao, Z.~Bi, and Y.~Wan, ``A joint power allocation and scheduling
  algorithm based on quasi-interference alignment in underwater acoustic
  networks,'' in \emph{OCEANS 2022-Chennai}.\hskip 1em plus 0.5em minus
  0.4em\relax IEEE, 2022, pp. 1--6.

\bibitem{gou2023achieving}
Y.~Gou, T.~Zhang, J.~Liu, T.~Yang, S.~Song, and J.-H. Cui, ``Achieving
  fair-effective communications and robustness in underwater acoustic sensor
  networks: A semi-cooperative approach,'' \emph{IEEE Transactions on Mobile
  Computing}, 2023.

\bibitem{mnih2015human}
V.~Mnih, K.~Kavukcuoglu, D.~Silver, A.~A. Rusu, J.~Veness, M.~G. Bellemare,
  A.~Graves, M.~Riedmiller, A.~K. Fidjeland, G.~Ostrovski \emph{et~al.},
  ``Human-level control through deep reinforcement learning,'' \emph{nature},
  vol. 518, no. 7540, pp. 529--533, 2015.

\bibitem{valerio2015reinforcement}
V.~D. Valerio, C.~Petrioli, L.~Pescosolido, and M.~Van Der~Shaar, ``A
  reinforcement learning-based data-link protocol for underwater acoustic
  communications,'' in \emph{Proceedings of the 10th International Conference
  on Underwater Networks \& Systems}, 2015, pp. 1--5.

\bibitem{wang2019self}
H.~Wang, Y.~Li, and J.~Qian, ``Self-adaptive resource allocation in underwater
  acoustic interference channel: A reinforcement learning approach,''
  \emph{IEEE Internet of Things Journal}, vol.~7, no.~4, pp. 2816--2827, 2019.

\bibitem{he2020trust}
Y.~He, G.~Han, J.~Jiang, H.~Wang, and M.~Martinez-Garcia, ``A trust update
  mechanism based on reinforcement learning in underwater acoustic sensor
  networks,'' \emph{IEEE Transactions on Mobile Computing}, vol.~21, no.~3, pp.
  811--821, 2022.

\bibitem{urick1983principles}
R.~J. Urick, ``Principles of underwater sound 3rd edition,'' \emph{Peninsula
  Publising Los Atlos, California}, vol.~22, pp. 23--24, 1983.

\bibitem{morozs2020channel}
N.~Morozs, W.~Gorma, B.~T. Henson, L.~Shen, P.~D. Mitchell, and Y.~V. Zakharov,
  ``Channel modeling for underwater acoustic network simulation,'' \emph{IEEE
  Access}, vol.~8, pp. 136\,151--136\,175, 2020.

\bibitem{stojanovic2007relationship}
M.~Stojanovic, ``On the relationship between capacity and distance in an
  underwater acoustic communication channel,'' \emph{ACM SIGMOBILE Mobile
  Computing and Communications Review}, vol.~11, no.~4, pp. 34--43, 2007.

\bibitem{etter2018underwater}
P.~C. Etter, \emph{Underwater acoustic modeling and simulation}.\hskip 1em plus
  0.5em minus 0.4em\relax CRC press, 2018.

\bibitem{dhanak2016springer}
M.~R. Dhanak and N.~I. Xiros, \emph{Springer handbook of ocean
  engineering}.\hskip 1em plus 0.5em minus 0.4em\relax Springer, 2016.

\bibitem{pompili2009cdma}
D.~Pompili, T.~Melodia, and I.~F. Akyildiz, ``A cdma-based medium access
  control for underwater acoustic sensor networks,'' \emph{IEEE Transactions on
  Wireless Communications}, vol.~8, no.~4, pp. 1899--1909, 2009.

\bibitem{song2019optimizing}
Y.~Song and P.-Y. Kong, ``Optimizing design and performance of underwater
  acoustic sensor networks with 3d topology,'' \emph{IEEE Transactions on
  Mobile Computing}, vol.~19, no.~7, pp. 1689--1701, 2019.

\bibitem{zhu2014toward}
Y.~Zhu, Z.~Peng, J.-H. Cui, and H.~Chen, ``Toward practical mac design for
  underwater acoustic networks,'' \emph{IEEE Transactions on Mobile Computing},
  vol.~14, no.~4, pp. 872--886, 2014.

\bibitem{zhao2022coach}
J.~Zhao, Y.~Zhao, W.~Wang, M.~Yang, X.~Hu, W.~Zhou, J.~Hao, and H.~Li,
  ``Coach-assisted multi-agent reinforcement learning framework for unexpected
  crashed agents,'' \emph{Frontiers of Information Technology \& Electronic
  Engineering}, vol.~23, no.~7, pp. 1032--1042, 2022.

\bibitem{wei2016power}
L.~Wei, Z.~Wang, J.~Liu, Z.~Peng, and J.-H. Cui, ``Power efficient deployment
  planning for wireless oceanographic systems,'' \emph{IEEE Systems Journal},
  vol.~12, no.~1, pp. 516--526, 2016.

\bibitem{islam2022lifetime}
K.~Y. Islam, I.~Ahmad, D.~Habibi, J.~Jin, and M.~Waqas, ``Lifetime maximization
  in underwater wireless communication networks,'' \emph{IEEE Sensors Journal},
  vol.~22, no.~15, pp. 15\,549--15\,560, 2022.

\bibitem{stamatiou2011throughput}
K.~Stamatiou, P.~Casari, and M.~Zorzi, ``Throughput and transmission capacity
  of underwater networks with randomly distributed nodes,'' in \emph{2011 IEEE
  Global Telecommunications Conference-GLOBECOM 2011}.\hskip 1em plus 0.5em
  minus 0.4em\relax IEEE, 2011, pp. 1--5.

\bibitem{freitag2005whoi}
L.~Freitag, M.~Grund, S.~Singh, J.~Partan, P.~Koski, and K.~Ball, ``The whoi
  micro-modem: An acoustic communications and navigation system for multiple
  platforms,'' in \emph{Proceedings of OCEANS 2005 MTS/IEEE}.\hskip 1em plus
  0.5em minus 0.4em\relax IEEE, 2005, pp. 1086--1092.

\bibitem{diamant2018fair}
R.~Diamant, P.~Casari, F.~Campagnaro, O.~Kebkal, V.~Kebkal, and M.~Zorzi,
  ``Fair and throughput-optimal routing in multimodal underwater networks,''
  \emph{IEEE Transactions on Wireless Communications}, vol.~17, no.~3, pp.
  1738--1754, 2018.

\bibitem{diamant2013robust}
R.~Diamant, G.~N. Shirazi, and L.~Lampe, ``Robust spatial reuse scheduling in
  underwater acoustic communication networks,'' \emph{IEEE Journal of Oceanic
  Engineering}, vol.~39, no.~1, pp. 32--46, 2013.

\bibitem{jain2008art}
R.~Jain, \emph{The art of computer systems performance analysis}.\hskip 1em
  plus 0.5em minus 0.4em\relax john wiley \& sons, 2008.

\bibitem{foerster2018counterfactual}
J.~Foerster, G.~Farquhar, T.~Afouras, N.~Nardelli, and S.~Whiteson,
  ``Counterfactual multi-agent policy gradients,'' in \emph{Proceedings of the
  AAAI conference on artificial intelligence}, vol.~32, no.~1, 2018.

\bibitem{casari2020asuna}
P.~Casari, F.~Campagnaro, E.~Dubrovinskaya, R.~Francescon, A.~Dagan, S.~Dahan,
  M.~Zorzi, and R.~Diamant, ``Asuna: A topology data set for underwater network
  emulation,'' \emph{IEEE Journal of Oceanic Engineering}, vol.~46, no.~1, pp.
  307--318, 2020.

\bibitem{hausknecht2015deep}
M.~Hausknecht and P.~Stone, ``Deep recurrent q-learning for partially
  observable mdps,'' in \emph{2015 aaai fall symposium series}, 2015.

\bibitem{liu2018td}
J.~Liu, T.~Zhang, G.~Han, and Y.~Gou, ``Td-lstm: Temporal dependence-based lstm
  networks for marine temperature prediction,'' \emph{Sensors}, vol.~18,
  no.~11, p. 3797, 2018.

\bibitem{xia2021multi}
Z.~Xia, J.~Du, J.~Wang, C.~Jiang, Y.~Ren, G.~Li, and Z.~Han, ``Multi-agent
  reinforcement learning aided intelligent uav swarm for target tracking,''
  \emph{IEEE Transactions on Vehicular Technology}, vol.~71, no.~1, pp.
  931--945, 2021.

\bibitem{sunehag2018value}
P.~Sunehag, G.~Lever, A.~Gruslys, W.~M. Czarnecki, V.~F. Zambaldi,
  M.~Jaderberg, M.~Lanctot, N.~Sonnerat, J.~Z. Leibo, K.~Tuyls \emph{et~al.},
  ``Value-decomposition networks for cooperative multi-agent learning based on
  team reward,'' in \emph{AAMAS}, 2018.

\bibitem{zhao2022deep}
N.~Zhao, N.~Yao, Z.~Gao, and Z.~Lu, ``Deep reinforcement learning based
  time-domain interference alignment scheduling for underwater acoustic
  networks,'' \emph{Journal of Marine Science and Engineering}, vol.~10, no.~7,
  p. 903, 2022.

\bibitem{ye2021scalable}
Y.~Ye, Y.~Tang, H.~Wang, X.-P. Zhang, and G.~Strbac, ``A scalable
  privacy-preserving multi-agent deep reinforcement learning approach for
  large-scale peer-to-peer transactive energy trading,'' \emph{IEEE
  transactions on smart grid}, vol.~12, no.~6, pp. 5185--5200, 2021.

\bibitem{li2020routing}
X.~Li, X.~Hu, R.~Zhang, and L.~Yang, ``Routing protocol design for underwater
  optical wireless sensor networks: A multiagent reinforcement learning
  approach,'' \emph{IEEE Internet of Things Journal}, vol.~7, no.~10, pp.
  9805--9818, 2020.

\bibitem{hu2010qelar}
T.~Hu and Y.~Fei, ``Qelar: A machine-learning-based adaptive routing protocol
  for energy-efficient and lifetime-extended underwater sensor networks,''
  \emph{IEEE transactions on mobile computing}, vol.~9, no.~6, pp. 796--809,
  2010.

\bibitem{zhao2020sim}
W.~Zhao, J.~P. Queralta, and T.~Westerlund, ``Sim-to-real transfer in deep
  reinforcement learning for robotics: a survey,'' in \emph{2020 IEEE symposium
  series on computational intelligence (SSCI)}.\hskip 1em plus 0.5em minus
  0.4em\relax IEEE, 2020, pp. 737--744.

\bibitem{arlpy2023}
\BIBentryALTinterwordspacing
M.~Chitre. (2018) Arl python tools. [Online]. Available:
  \url{https://pypi.org/project/arlpy/}
\BIBentrySTDinterwordspacing

\bibitem{lee2011distributed}
H.-W. Lee, E.~Modiano, and L.~B. Le, ``Distributed throughput maximization in
  wireless networks via random power allocation,'' \emph{IEEE transactions on
  mobile computing}, vol.~11, no.~4, pp. 577--590, 2011.

\end{thebibliography}

%

\vspace{-1em}
\begin{IEEEbiography}[{\includegraphics[width=1in,height=1.25in,clip,keepaspectratio]{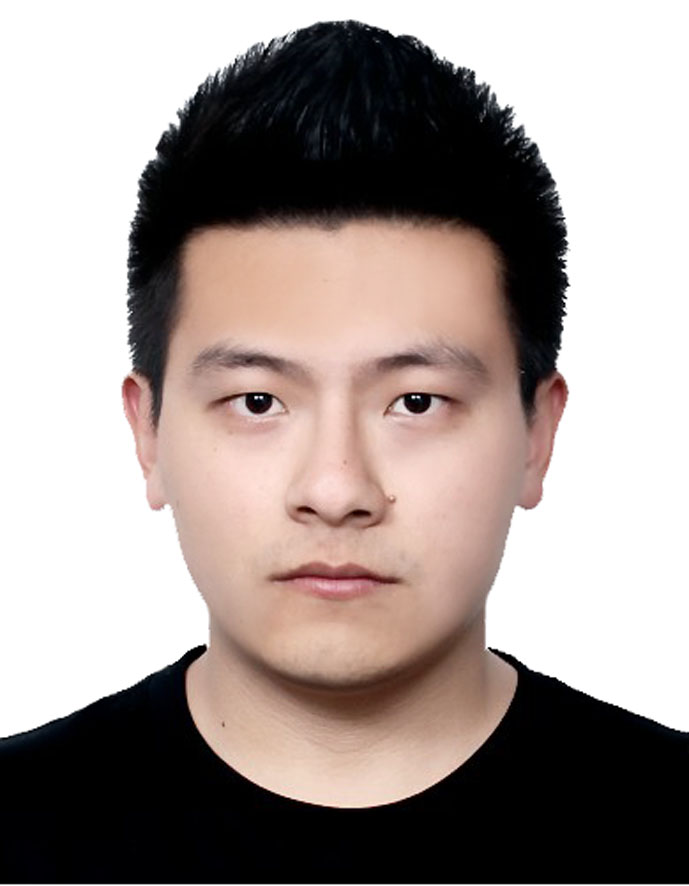}}]{Tong Zhang} received his B.S. degree (2014) from the College of Mathematics and Computer Science, Fuzhou University, Fuzhou, China, M.S. degree (2018) and PhD degree (2023) from the College of Computer Science and Technology, Jilin University, Changchun, China. He is currently a Post Doctoral Researcher with Beihang Ningbo Innovation Research Institute, Beihang University, Ningbo, China. His research interests include the resource management of UWSNs and deep reinforcement learning.
\end{IEEEbiography}
\begin{IEEEbiography}[{\includegraphics[width=1in,height=1.25in,clip,keepaspectratio]{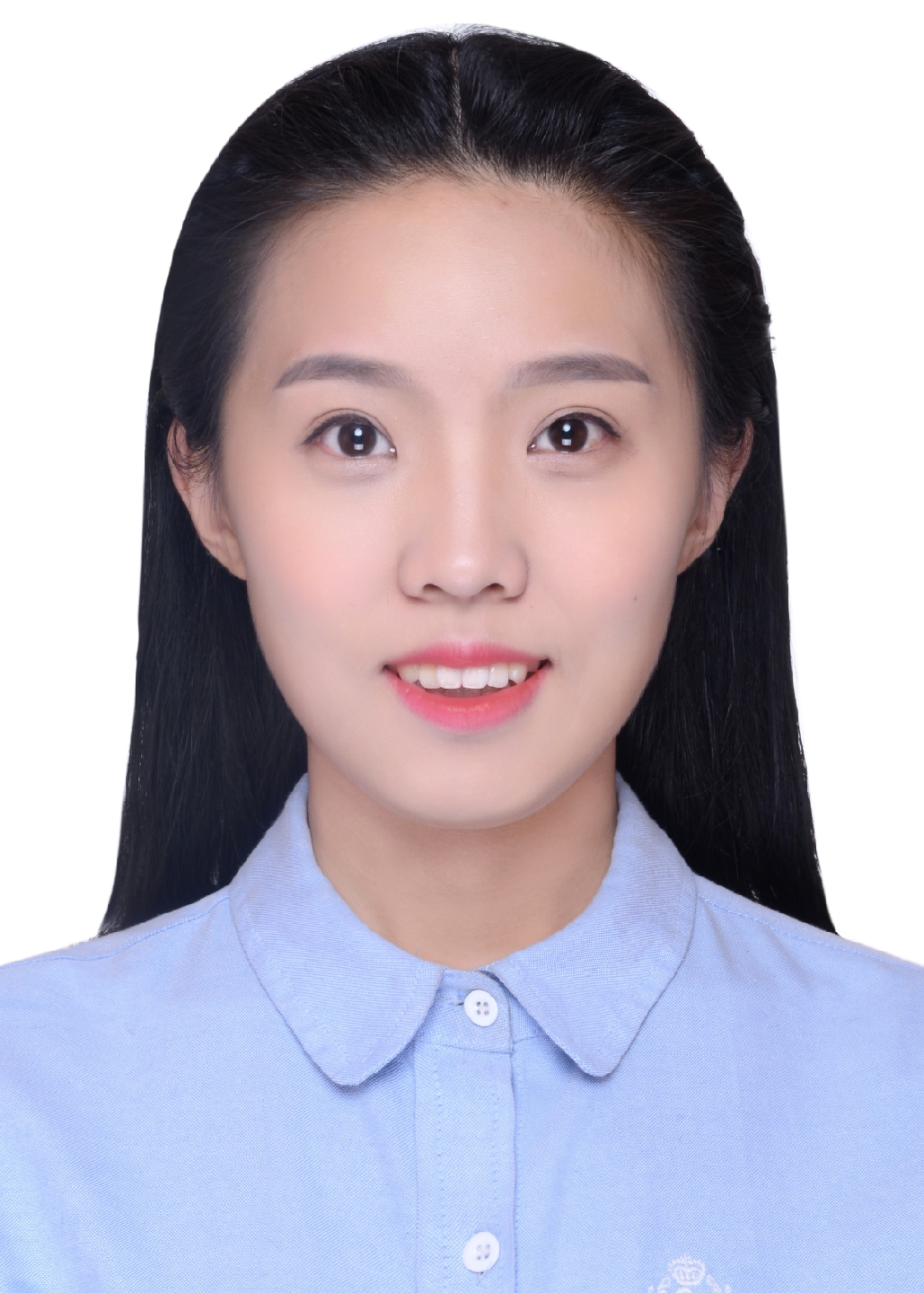}}]{Yu Gou} received her B.S. degree (2015), M.S. degree (2018), and PhD degree (2023) from the College of Computer Science and Technology, Jilin University, Changchun, China. She is currently a Post Doctoral Researcher with Beihang Ningbo Innovation Research Institute, Beihang University, Ningbo, China. Her research interests are include deep multi-agent reinforcement learning, underwater network performance optimization, and robust underwater networks.
\end{IEEEbiography}
\vspace{-1em}
\begin{IEEEbiography}[{\includegraphics[width=1in,height=1.25in,clip,keepaspectratio]{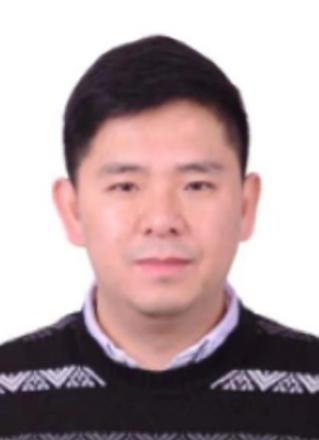}}]{Jun Liu} received the BEng degree (2002) in computer science from Wuhan University, China, the PhD degree (2013) in Computer Science and Engineering from University of Connecticut, USA. Currently, he is a professor of the School of Electronic and Information Engineering at Beihang University, China. His major research focuses on underwater networking, synchronization, and localization. He is a member of the IEEE Computer Society.
\end{IEEEbiography}
\begin{IEEEbiography}[{\includegraphics[width=1in,height=1.25in,clip,keepaspectratio]{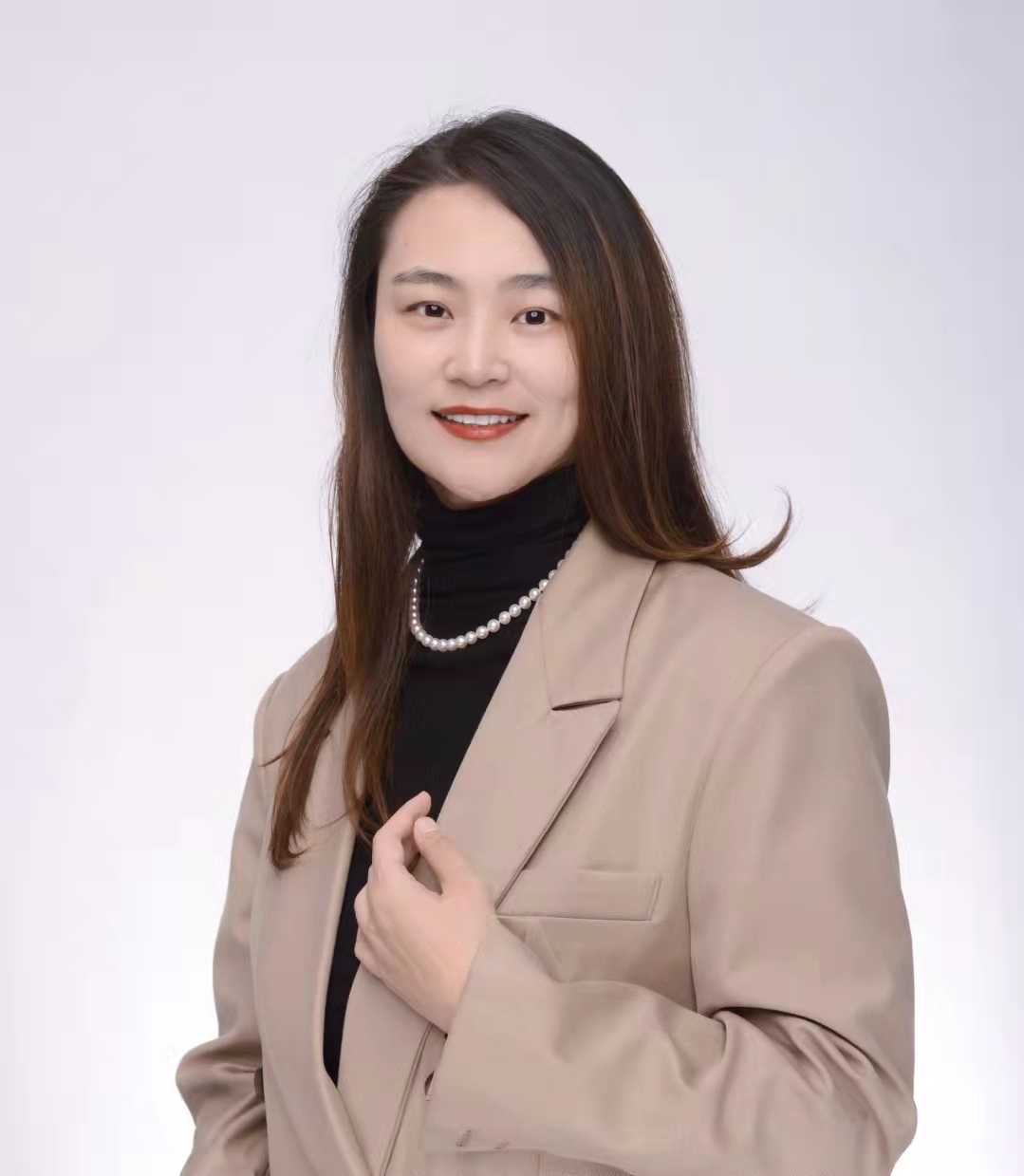}}]{Shanshan Song} received the BS degree (2011) and MS degree (2014) in computer science and technology from Jilin University, China, received PhD degree (2018) in Management science and engineering from Jilin University, China. Currently, she is the associate professor of the College of Computer Science and Technology at Jilin University, China. Her major research focuses on underwater  localization and navigation and machine learning.
\end{IEEEbiography}
\begin{IEEEbiography}[{\includegraphics[width=1in,height=1.25in,clip,keepaspectratio]{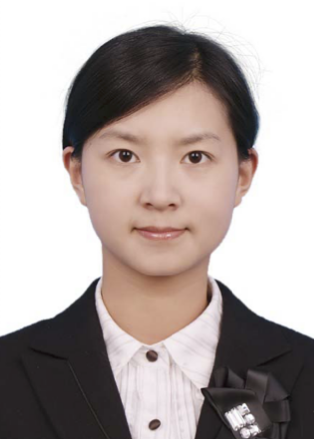}}]{Tingting Yang} (M’13) received her B.Sc. (2004) and Ph.D. degrees (2010) from Dalian Maritime University, China. She is currently a Professor at Dalian Maritime University, and also at Pengcheng Laboratory. Her research interests are in the areas of space-air-ground-sea integrated networks, and network AI. She serves as the associate Editor-in-Chief of the IET Communications, as well as the advisory editor for SpringerPlus.
\end{IEEEbiography}
\begin{IEEEbiography}[{\includegraphics[width=1in,height=1.25in,clip,keepaspectratio]{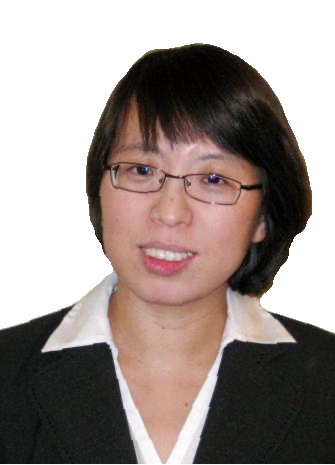}}]{Jun-Hong Cui} received the BS degree (1995) in computer science from Jilin University, China, the MS degree (1998) in computer engineering from Chinese Academy of Sciences, and the PhD degree (2003) in computer science from UCLA. Currently, she is on the faculty of Shenzhen Institute for Advanced Study, UESTC, Shenzhen, China. Recently, her research mainly focuses on algorithm and protocol design in UWSNs.
\end{IEEEbiography}




\end{document}